\documentclass[%
 reprint,
groupedaddress,
 amsmath,amssymb,
 aps,
pra,
]{revtex4-2}
\usepackage{graphicx}
\usepackage{xcolor}
\usepackage{dcolumn}
\usepackage{bm}
\usepackage{physics}
\usepackage{amsthm}
\usepackage{dsfont}
\usepackage{url}
\usepackage{hyperref}
\usepackage{booktabs}
\newtheorem{theorem}{Theorem}
 
\newtheorem{proposition}{Proposition} 
\newtheorem{lemma}[theorem]{Lemma}

\newcommand{\GHZket}{\ket{\mathrm{GHZ}_{\text{qubit}}}}

\newcommand{\GHZketdrei}{\ket{\mathrm{GHZ}_{\text{qutrit}}}}
\newcommand{\GHZbradrei}{\bra{\mathrm{GHZ}_{\text{qutrit}}}}

\newcommand{\GF}{\mathbb{F}_2}
\newcommand{\GFn}{\mathbb{F}_2^n}

\newcommand{\GFp}{\mathbb{F}_p}

\newcommand{\GFt}{\mathbb{F}_3}
\newcommand{\GFpn}{\mathbb{F}_p^n}
\newcommand{\GFtn}{\mathbb{F}_3^n}

\newcommand{\GFpl}{\mathbb{F}_p^l}

\newcommand{\ps}{\bar{p}_{S}}

\newcommand{\psc}{\bar{p}_S^c}
\newcommand{\psq}{\bar{p}_S^q}

\begin{document}

\title{The power of qutrits for non-adaptive measurement-based quantum computing}

\author{Jelena Mackeprang$^{1,2}$, Daniel Bhatti$^{1,2}$, Matty J. Hoban$^{3,*}$, Stefanie Barz$^{1,2}$}
\affiliation{
$^{1}$Institute for Functional Matter and Quantum Technologies, University of Stuttgart, 70569 Stuttgart, Germany\\
$^{2}$Center for Integrated Quantum Science and Technology (IQST), University of Stuttgart, 70569 Stuttgart, Germany\\
$^{3}$Department of Computing, Goldsmiths, University of London, New Cross, London SE14 6NW, United Kingdom\\
$^{*}$Present address: Quantinuum and Cambridge Quantum Computing Ltd, London SW1P 1BX, United Kingdom}

\begin{abstract}
Non-locality is not only one of the most prominent quantum features but can also serve as a resource for various information-theoretical tasks. Analysing it from an information-theoretical perspective has linked it to applications such as non-adaptive measurement-based quantum computing (NMQC). In this type of quantum computing the goal is to output a multivariate function. The success of such a computation can be related to the violation of a generalised Bell inequality. So far, the investigation of binary NMQC with qubits has shown that quantum correlations can compute all Boolean functions using at most $2^n-1$ qubits, whereas local hidden variables (LHVs) are restricted to linear functions. Here, we extend these results to NMQC with qutrits and prove that quantum correlations enable the computation of all ternary functions using the generalised qutrit Greenberger-Horne-Zeilinger (GHZ) state as a resource and at most $3^n-1$ qutrits. This yields a corresponding generalised GHZ type paradox for any ternary function that LHVs cannot compute. We give an example for an $n$-variate function that can be computed with only $n+1$ qutrits, which leads to convenient generalised qutrit Bell inequalities whose quantum bound is maximal. Finally, we prove that not all functions can be computed efficiently with qutrit NMQC by presenting a counterexample.
\end{abstract}

\maketitle

\section{\label{sec:intor} Introduction}

Measurement-based quantum computing uses adaptive single-qubit measurements on highly entangled resource states, together with feedforward, to implement universal quantum computing~\cite{Raussendorf2001}. 
Achieving this adaptivity experimentally requires fast switching of measurement bases, which is technically very challenging.

However, even a reduced measurement-based model without adaptivity---non-adaptive measurement-based quantum computing (NMQC)---allows universal classical computing~\cite{Anders2009,Hoban2011IOP}.
Beyond that, NMQC opens up new avenues for studying fundamental questions in quantum physics, in particular, non-locality~\cite{Cleve2004}.
One can show that quantum correlations have an advantage over local hidden variables (LHVs) in the computation of classical functions, and the success probability of computing a function correctly has been linked to the violation of a generalised Bell inequality~\cite{Hoban2011IOP}.
The fact that in NMQC the distinct measurement sites do not communicate enables a space-like separation, which is needed for loophole-free Bell test experiments.

So far, NMQC has primarily been discussed in the context of computing Boolean functions and qubit resource states. Though few generalisations to higher-dimensional functions have been considered, the discussion mostly revolved around the limitations of LHVs~\cite{Hoban2011PRA,Frembs2018}. The full power of quantum correlations of higher-dimensional systems, qudits, has remained largely unexplored.

In the broader field of circuit-based quantum computing, qudits have been subject to increased research interest in recent years~\cite{Wang2020}, and some generalisations of adaptive measurement-based quantum computing to higher dimensional systems have already been discussed~\cite{Zhou2003,Booth2021}. The denser encoding of information, and thus the reduction of computational costs compared to qubits, as well as the possible simplification of experiments makes them a good candidate for next-generation quantum computing~\cite{Wang2020,Low2020,Blok2021,Hill2021}. In particular, three-dimensional quantum systems, qutrits, have been studied~\cite{Klimov2003,Randall2015,Gokhale2020,Yurtalan2020,Low2020,Hill2021,Wu2021,Blok2021}. 
Various recent experiments have explored qutrits for quantum computing and quantum information utilising superconducting~\cite{Pakhomchik2020,Blok2021,Nikolaeva2021,Galda2021,Cervera-Lierta2021} or photonic systems~\cite{Schlederer2016,Babazadeh2017,Borges2018,Erhard2018,Luo2019,Hu2020}, and ion-trap quantum processors~\cite{Ringbauer2021}.

In this work, we investigate the power of qutrits and their quantum correlations in the framework of NMQC. To distinguish NMQC with qutrits from its qubit analogue, we will use the term \textit{3-NMQC} in contrast to \textit{2-NMQC} for qubits.

We study the generalised $l$-qutrit GHZ state
\begin{equation}\label{eq:3GHZ}
	\GHZketdrei \equiv \frac{1}{\sqrt{3}}\left(\ket{0}^{\otimes l}+ \ket{1}^{\otimes l}+\ket{2}^{\otimes l}\right) ,
\end{equation}
as a resource state for 3-NMQC and show that it enables the computation of all ternary functions $f:\{0,1,2\}^n\mapsto \{0,1,2\}$ using at most $l=3^n-1$ qutrits. 
In particular, any such ternary function induces a generalised qutrit Bell inequality for which the quantum violation is maximal. 
To show the power of 3-NMQC we give an example for a family of $n$-variate functions whose computation requires only $n+1$ qutrits. Furthermore, we prove that despite enabling the computation of all ternary functions, it is not possible to compute all of them efficiently with the qutrit GHZ state. Our work paves the way for measurement-based quantum computing with qudits that may have an advantage with regard to scalability and information encoding compared to their qubit counterparts.


%
\section{\label{sec:NMQC} NMQC}
%
We will first give an introduction into NMQC and its relation to Bell inequalities by summarising the most relevant results and literature.

%
\subsection{\label{sec:BG:pNMQC} General NMQC}
%
We start by introducing the concept of NMQC for $p$-dimensional quantum systems and $p$-valued logic, $p$ being prime. The goal is to compute a function $f: \GFpn \rightarrow \GFp$. $\GFp \equiv (\{0,1,\ldots,p-1\}, \oplus , \cdot)$ is the finite field of order $p$ consisting of the set $\{0,1,\ldots,p-1\}$ equipped with addition and multiplication $\mod p$, denoted by $\oplus$ and $\cdot$. $ \GFpn = (\{0,1, \ldots, p-1 \}^n, \oplus, \cdot)$) denotes the $n$-dimensional coordinate vector space over $\GFp$. \\
For the remainder of our work, $p$ will either be equal to \textit{two}, meaning  the procedure uses qubits and binary logic, or equal to \textit{three}, meaning qutrits and ternary logic. \newline
 The procedure goes as follows (see also Fig.~\ref{fig:NMQCcomp}).:

\begin{enumerate}

\item Pre-processing: The starting point of NMQC is always a restricted computer limited to addition $\mod p$.
For $p=2$, it operates on bits and computes parities, i.e addition mod $2$ in $\{0,1\}$. For $p=3$, it operates on trits and uses addition mod $3$ in $\{0,1,2\}$. 

This restricted computer now pre-processes an $n$-dimensional $p$-valued input vector $x$ $\in \GFpn$ and turns it into an $l$-dimensional $p$-valued output vector $s$ ($\in \GFpl$). This pre-processing procedure can be seen as a matrix-vector multiplication:
	\begin{equation} \label{eq:BG:sPx}
		s = (Px)_{\oplus},
	\end{equation}
where $P$ is an $l$-by-$n$ matrix with elements in $\GFp$ and the $\oplus$ in the index denotes that the matrix-vector product is evaluated with respect to $\GFp$~\cite{Hoban2011PRA}.

	\item Measurement settings: The elements of $s$ ($s_0, ...,s_{l-1}$) now determine the settings for measurements on a computational resource state. This resource state has $l$ qu-$p$-it subsystems, e.g. $l$ qubits or qutrits. 
	
	For each subsystem, we have a set of $p$ measurement operators $\hat{m}_i$, e.g. two different operators for qubit systems and three for qutrit systems.
	The number of measurement outcomes per measurement operator is also $p$, meaning qubit measurements have two possible outcomes and qutrit measurements have three possible outcomes.

	\item Measurement results: Each measurement yields a measurement result $M_i \in \Omega$. Here, $\Omega$ depends on $p$ and is defined as $\Omega = \{\omega^0,\omega^1,...,\omega^{p-1}\}$ with $\omega = e^{\frac{2\pi i}{p}}$. For $p=2$, the results are in $\{1, -1\}$, where for $p=3$, they are elements of $\{1, e^{\frac{2\pi i}{3}},  e^{\frac{4\pi i}{3}}\}$.

	\item Mapping: The values $M_i$ are now mapped to the values $m_i \in \GFp$ by:
	\begin{equation}\label{eq:M_i}
		M_i = \omega^{m_i}.
	\end{equation}
	This means, for qubits, we simply map $\{1, -1\}$ onto $\{0, 1\}$, whereas for qutrits, we map $\{1, e^{\frac{2\pi i}{3}},  e^{\frac{4\pi i}{3}}\}$ onto $\{0, 1,2\}$ .

	\item Post-processing: We collect all $m_i$ in a vector $m$ that now contains the $l$ measurement results. This vector is then sent back to the pre-processor to compute the sum over all of the $m_i$, yielding a value $z \in \GFp$:
	\begin{equation}\label{eq:z}
		z \equiv \bigoplus_{i=0}^{l-1} m_i,
	\end{equation}
	
	\item Verification: The computation (for a given input vector $x$) is successful if $f(x)=z$. We say that an NMQC scheme is \textbf{deterministic} if it outputs $z = f(x)$ for every $x$.  

\end{enumerate}


\subsection{\label{sec:BG:2NMQC}2-NMQC}


For 2-NMQC it has been shown that if the measurement statistics are described by LHVs \cite{Bell1964},  the output $z$ is restricted to linear functions, i.e. functions that can be written as follows:
\begin{equation} \label{eq:affine}
	f(x) = \bigoplus_{i=0}^{n-1} v_i \cdot x_i \oplus c, \quad v \in \GFn, c \in \GF.
\end{equation}

The pre-processing computer is already capable of outputting linear functions, i.e. functions that can be written as $f(x)=\bigoplus_{i=0}^{n-1} v_i x_i$. The additional bit $c$ can be added in post-processing. This means that LHVs do not ``boost'' the pre-processor in any way~\cite{Hoban2011IOP}. 
Note that, we use the term \textit{linear} for functions of the form $f(x)=\bigoplus_{i=0}^{n-1} v_ix_i\oplus c$, $c \in {1,2}$, and \textit{strictly linear} for those of the form $f(x)=\bigoplus_{i=0}^{n-1} =v_ix_i$.

Non-local quantum correlations, however, can enhance the pre-processor to classical universality. Using the generalised $l$-qubit GHZ state

\begin{equation} \label{eq:2GHZ}
	\GHZket= \frac{1}{\sqrt{2}}\left(\ket{0}^{\otimes l}+\ket{1}^{\otimes l}\right),	\end{equation}
one can compute \emph{all} functions $f: \GFn \rightarrow \GF$ with at most $l=2^n-1$ qubits. The computation of a non-linear function---one that cannot be written as in Eqn.~\eqref{eq:affine}---thus requires \emph{non-locality} and can be seen as a type of GHZ paradox~\cite{Hoban2011IOP}. \\

One can show that the $n$-bit pairwise AND function
\begin{equation} \label{eq:pairwiseand}
	g_n(x) \equiv \bigoplus_{j=0}^{n-2}\left(x_j\bigoplus_{k=j+1}^{n-1}x_k\right),
\end{equation}
can be computed efficiently with only $l=n+1$ qubits prepared in the state given by Eqn.~\eqref{eq:2GHZ}~\cite{Hoban2011IOP}. 
In contrast, the computation of the $n$-tuple AND function
\begin{equation}\label{eq:andn}
	\mathrm{AND}_n(x) \equiv \prod_{i=0}^{n-1} x_i
\end{equation}
demands at least $2^n-1$ qubits~\cite{Hoban2011IOP}. Therefore, not every Boolean function can be \emph{efficiently} computed with 2-NMQC. Note that the pairwise AND function given by Eqn.~\eqref{eq:pairwiseand} is invariant with respect to coordinate permutation. The $n$-tuple AND function given by Eqn.~\eqref{eq:andn} \emph{only outputs 1} if \emph{all} input bits are 1 and outputs 0 in all other cases. We will use these two observations for our own results in section~\ref{sec:ineff}.  


\subsection{2-NMQC and Bell inequalities} \label{sec:NMQCBell}


The success of a 2-NMQC computation is closely related to the violation of a Bell inequality. 
In particular, every Boolean function $f: \GFn \rightarrow \GF$ induces a 2-NMQC game. 
Further, each game is defined by a sampling probability distribution $w(x)$ and a pre-processing $P$.
The average success probability of such a game is bounded by a Bell inequality.

In each round of the game, an input $x \in \GFn$ is sampled with respect to $w(x)$ before the entire NMQC procedure from above is performed. If the output $z=f(x)$, the round is won. The pre-processing and the resource state must thus be chosen, so that the total average success probability is maximised. This average success probability $\ps$ is defined by \cite{Hoban2011IOP}:
\begin{equation} \label{eq:psdef}
	\ps = \sum_{x} w(x)p(z=f(x)|s=(Px)_{\oplus}),
\end{equation}
where $p(z=f(x)|s=(Px)_{\oplus})$ is the probability that $z$ is equal to $f(x)$ given that the measurement settings are $s=(Px)_{\oplus}$.  \\

It can be shown that $\ps$ is related to the bound $u$ of a Bell inequality in terms of expectation values~\cite{Hoban2011IOP}:
\begin{align} 
	\ps &=  \frac{1+u}{2} \label{eq:psu},\\
	u &= \sum_{x} (-1)^{f(x)}w(x) E(z|s) \leq \begin{cases}  c \\ q 
	\end{cases}.
	\label{eq:psu2}
\end{align}

The inequality~\eqref{eq:psu2} is a normalised Bell inequality with a classical (LHV) bound $c$ and a quantum bound $q$. 
The expectation values are defined to be:
\begin{equation}\label{eq:E2def}
	E(z|s) = p(z=0|s)-p(z=1|s),
\end{equation}
where $p(z=0|s)$ is the probability that $z$ is equal to $0$ and $p(z=1|s)$ is the probability that $z$ is equal to $1$ given the measurement settings were determined by $s$.

Note that deterministic NMQC for an $n$-variate function corresponds to a probabilistic NMQC game where $w(x)= 1/2^n$ with the quantum average success probability $\psq=1$, and thus $q=1$.

Eqn.~\eqref{eq:psu2} is an element of the complete set of $n$-variate Bell inequalities with two possible measurement settings and two possible measurement results per site, called $(n,2,2)$ Bell inequalities. As the term $(-1)^{f(x)}w(x)$ can realise any real number between -1 and 1, the \emph{entire set} of (normalised) $(n,2,2)$ Bell inequalities can be found through probabilistic 2-NMQC games~\cite{Hoban2011IOP}. 
The GHZ state always maximally violates the $(n,2,2)$ Bell inequalities and minimises the number of required qubits for a violation~\cite{Werner2001,Zukowski2002}. It is thus also \emph{optimal} for both probabilistic and deterministic 2-NMQC~\cite{Hoban2011IOP}. 

Next, we briefly discuss what is already known about 3-NMQC.


\subsection{\label{sec:pNMQC}3-NMQC} \label{sec:3NMQC}

\begin{figure*}[ht]
		\centering
			\includegraphics[width=.9\linewidth]{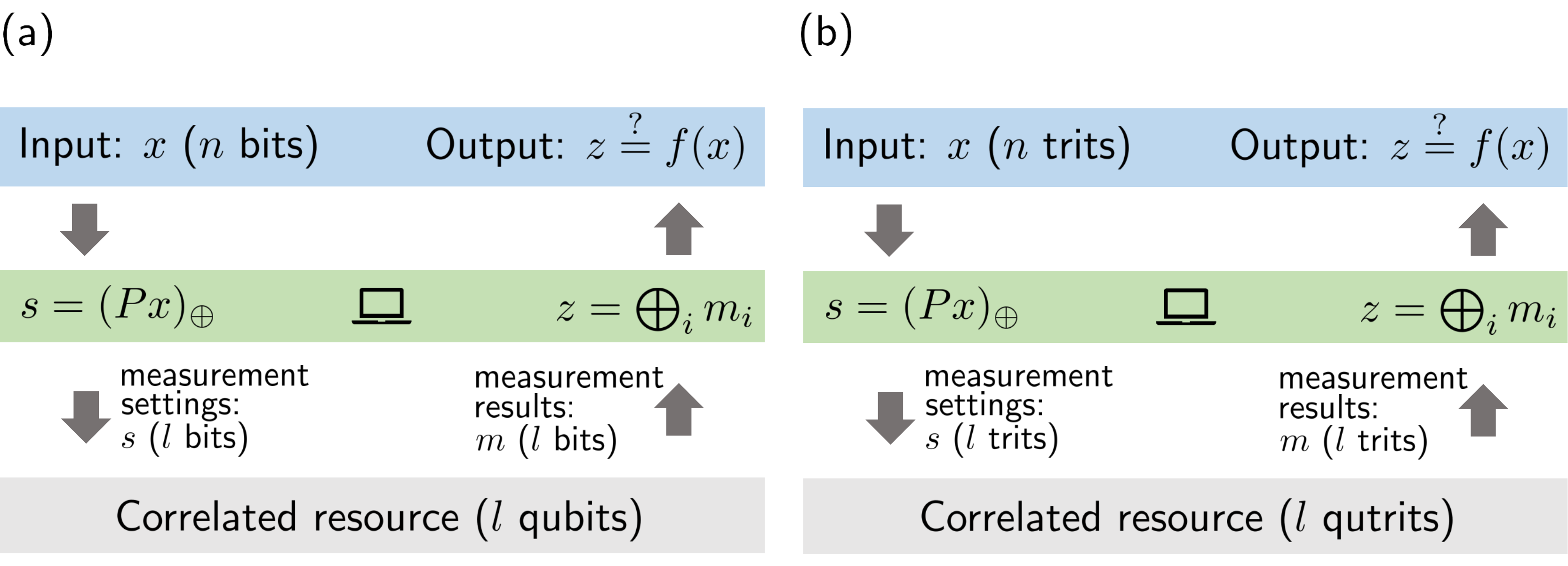}
				\caption{Schematic representation of the 2-NMQC and 3-NMQC procedures. In both cases, an input vector $x$ with $n$ elements is sent to the pre-processor. For 2-NMQC, these elements are bits (elements of $\{0,1\}$) whereas for 3-NMQC, these elements are trits (elements of $\{0,1,2\}$).
		The pre-processor takes this $n$-entry vector and transforms it into another $l$-entry vector, called $s$, via $s=(Px)_{\oplus}$. 
		Note that the pre-processor operates with bits using mod 2 addition in 2-NMQC, while it operates with trits using mod 3 addition in 3-NMQC.
				The vector $s$ now contains information about the measurement settings on the computational resource, either consisting of qubits or qutrits. 
				The respective measurements at the $l$ measurement sites of the correlated resource are carried out and the measurement results are mapped to the values $m_i$.
				These values $m_i$ are summarized in the vector $m$, which is sent back to the pre-processor. The pre-processor then computes the sum over all $m_i$, called $z$, which is supposed to be equal to $f(x)$, either a Boolean function for 2-NMQC or a ternary function for 3-NMQC. }
				\label{fig:NMQCcomp}
	\end{figure*}

3-NMQC follows the NMQC procedure presented in Sec.~\ref{sec:BG:pNMQC} and shown in Fig.~\ref{fig:NMQCcomp}. For $p=3$, the possible measurement results are complex numbers and elements of $\Omega=\{1, e^{\frac{2\pi i}{3}},  e^{\frac{4\pi i}{3}}\}$.
Therefore, we require the measurement operators to be \emph{unitary} observables with eigenvalues being elements of $\Omega$. These unitary observables correspond to projective measurements whose outcomes are labelled by complex values~\cite{Hoban2011PRA,Lim2010, Arnault2012, Lawrence2017, Cervera-Lierta2019}~\footnote{Note that any unitary observable can be written as a sum with complex coefficients of two commuting Hermitian operators. So a measurement of a unitary observable can also be interpreted as the simultaneous measurement of two commuting Hermitian operators and the subsequent summation of their measurement results each multiplied by the appropriate complex coefficient.}.

LHVs in 3-NMQC are slightly boosted by the pre-processor~\cite{Hoban2011PRA}. Before the pre-processing, they are capable of computing functions $f: \GFtn \rightarrow \GFt$ that can be written as:
\begin{equation}
	f(x) = \bigoplus_{i=0}^{n-1} \bigoplus_{j=1}^2 (c_{i}^{(j)} x_i^j) \oplus c,
\end{equation}
where $c_{i}^{(j)} \in \GFt$ are coefficients and $c \in \GFt$ is an offset. After the pre-processing, LHVs can compute \emph{all polynomials up to a degree of} 2 \cite{Hoban2011PRA,Frembs2018}, meaning that they are slightly elevated by the pre-processing. Nevertheless, they are still incapable of computing all functions, as the maximum degree of an $n$-variable function over $\GFt$ is $n \cdot 2$. For example, LHVs cannot compute $f(x)=x_0^2x_1$. \newline
The computation of polynomials of a higher degree is a demonstration of non-locality \cite{Hoban2011PRA}, or more specifically \emph{strong} non-locality \cite{Frembs2018}, following the definition of strong contextuality and non-locality in \cite{Abramsky2011}. The differences between 2-NMQC and 3-NMQC are compared in Fig.~\ref{fig:NMQCcomp}. 

\subsection{3-NMQC and qutrit Bell inequalities} \label{sec:3NMQCBell}


Similar to 2-NMQC, there is a corresponding probabilistic 3-NMQC game for a given pre-processing $P$ and a given function, where in every round, an input $x \in \GFtn$ is sampled with respect to a sampling distribution $w(x)$. Its average success probability $\ps$ is analogously defined to the success probability of the probabilistic 2-NMQC game in Eqn.~\eqref{eq:psdef}: 
\begin{equation} \label{eq:psdef3}
	\ps = \sum_{x \in \GFtn} w(x)p(z=f(x)|s).
\end{equation}
$\ps$ can be related to a generalised Bell inequality via the discrete Fourier transform~\cite{Hoban2011PRA}:
\begin{equation} \label{eq:psBellqutrit}
	\ps =  \frac{1}{3} \sum_{x}  w(x)\sum_{M=0}^{2} \omega^{-f(x)M}E^{M}(z|s),
\end{equation} 
where $E^{M}(z|s)$ is defined as:
\begin{equation}
	\label{eq:EM}
	E^{M}(z|s) = \sum_{k=0}^{2} \omega^{kM} p(z=k|s).
\end{equation}

The right hand side of Eqn.~\eqref{eq:psBellqutrit} is a normalised sum over expectation values and can thus be seen as a generalised qutrit Bell inequality. Note that if $f(x) = z\; \forall x$, then the inner sum in Eqn.~\eqref{eq:psBellqutrit} will always be:
\begin{align}
\sum_{M=0}^{2} \omega^{-f(x)M}E^{M}(z|s) 
&=\sum_{M=0}^{2} \omega^{-f(x)M} \sum_{k=0}^{2} \omega^{kM} p(z=k|s) \nonumber \\
&= \sum_{M=0}^{2} \omega^{-f(x)M}  \omega^{f(x)M} \nonumber \\
&=3,
\end{align}
resulting in a success probability $\ps$ of 1. \newline
As is the case for qubits, the probabilistic 3-NMQC game can be interpreted as a way to demonstrate non-locality. The LHV success probability $\psc$ is limited by the LHV bound of the generalised Bell inequality on the right hand side of Eqn.~\eqref{eq:psBellqutrit}. A success probability that is larger than $\psc$ requires the violation of such a generalised Bell inequality and thus non-locality.

%
\section{Simplified Bell inequalities for 3-NMQC}\label{sec:3NMQCBellalt}
%

Having covered the necessary background, we continue with our own derivations. 
We start by simplifying the qutrit Bell inequalities and rewrite Eqn.~\eqref{eq:psBellqutrit}:
\begin{align} 
 \ps &=  \frac{1}{3} \sum_{x}  w(x)(1+\sum_{M=1}^{2} \omega^{-f(x)M}E^{M}(z|s)) \\
	&= \frac{1}{3}(1+\sum_{x}  w(x) \sum_{M=1}^{2} \omega^{-f(x)M}E^{M}(z|s)) \label{eq:psBell2}\\
		&\equiv \frac{1}{3} (1+ 2 u_3), \label{eq:psBellMatty}
 \end{align} 
where we have defined $u_3$ as the normalised sum over $x$ and $M$  in Eqn.~\eqref{eq:psBell2}.
As before, $c$ and $q$ are the LHV and the quantum bound of $u_3$. We can now interpret $u_3$ as a generalised Bell inequality bounded by $c$ for LHV and $q$ for quantum correlations:
 \begin{equation} 
 	 u_3 \leq \begin{cases} c \\ q  \end{cases}.
 \end{equation}

In Eqn.~\eqref{eq:psBell2} the sum runs from $M=1$ to 2, requiring the measurement of $3^n \times 2$ expectation values for an $n$-variate function. However, one can reduce this number of expectation values to be evaluated by exploiting the fact that  $\Re{e^{\frac{2\pi i}{3}}} = \Re{e^{\frac{4\pi i}{3}}} = -1/2$, leading to an alternative generalised Bell inequality: 
 \begin{equation}\label{eq:uprealmain}
\Re{u_3} = \Re{\sum_{x} w(x) \omega^{-f(x)}E(z|s)} \leq \begin{cases} c \\ q \end{cases},   
\end{equation}
which is related to $\ps$ via (see Appendix~\ref{sec:AppBell}):
\begin{equation} \label{eq:psBell3main}
	\ps = \frac{1}{3} (1+ 2 \Re{u_3}).
\end{equation}

Eqn.~\eqref{eq:uprealmain} requires only $3^n$ expectation values $E(z|s)$ in contrast to twice as many in  Eqn.~\eqref{eq:psBellqutrit}.
In the following, we base our derivations on the version of the generalised Bell inequality given by Eqn.~\eqref{eq:uprealmain}.

Note that the deterministic computation of a function $f: \GFtn \rightarrow \GFt$ with qutrits corresponds to a probabilistic 3-NMQC game with a uniform sampling distribution $w(x)=1/3^n$ where $q=1$.

\section{\label{sec:ghzuniversal} Quantum correlations elevate a ternary restricted computer to classical universality}


We now show that quantum correlations are capable of computing all ternary functions in 3-NMQC, as summarised in the following Theorem:

 \begin{theorem}\label{thm:ghzuniversal}
	One can deterministically compute all $n$-variate functions over $\GFt$ with 3-NMQC and $l$ qutrits using the generalised $l$-qutrit GHZ state $\GHZketdrei$, where $l$ is at most $3^n-1$.  
\end{theorem}

It follows directly from Eqn.~\eqref{eq:psBell3main} that any function $f: \GFtn \rightarrow \GFt$ induces a qutrit Bell inequality for which quantum correlations reach the maximum possible bound $q=1$. One can additionally conclude that for any ternary function that LHVs cannot compute, one can find an appropriate 3-NMQC procedure which can be seen as a generalised GHZ paradox.

 In the rest of this section, we sketch the main idea of the proof of Theorem~\ref{thm:ghzuniversal}. The full proof can be found in Appendix~\ref{sec:Appthm}. \\

%
\subsection{Main idea behind the proof of Theorem~\ref{thm:ghzuniversal}} \label{sec:proofsketch}
%
Due to the definition of 3-NMQC, we require unitary observables with eigenvalues in $\Omega$.

We choose the generalised qutrit GHZ state from Eqn.~\eqref{eq:3GHZ} as a resource and rotated generalised qutrit $X$ operators as measurement operators, similar to the qubit case in~\cite{Hoban2011IOP}.
The generalised $X$ and $Z$ operators for 3-dimensional quantum systems are given by  the operator $X$ that acts the following way on a qutrit in the state $\ket{n}$ (in the computational basis $\{\ket{0},\ket{1},\ket{2}\}$) \cite{Gottesman1999}: 
\begin{equation}
X\ket{n}=\ket{(n+1) \mod 3},
\end{equation}
and a $Z$ operator that acts the following way on a qutrit in the state $\ket{n}$ \cite{Gottesman1999}:
\begin{equation}
Z\ket{n} = \omega^n \ket{n},
\end{equation}
 The eigenvalues of both $X$ and $Z$ are $\Omega= \{1, \omega, \omega^2\}$ and the eigenstates of $Z$ are the computational basis states. \newline  
We can now define a rotated $X$ operator, leading to the measurement operator $\hat{m}_i$ corresponding to the $i$th element of $s$:
\begin{align}\label{eq:mi}
	\begin{split}
		\hat{m}_i &= Z^{\frac{s_i\phi_i}{3}}X Z^{-\frac{s_i\phi_i}{3}}\\
		&=\alpha^{s_i\phi_i}\ket{1}\bra{0}+\alpha^{s_i\phi_i}\ket{2}\bra{1}+\alpha^{-2 s_i \phi_i} \ket{0}\bra{2},
	\end{split}
\end{align}
where $\alpha = \omega^{\frac{1}{3}}$~\cite{Lawrence2017,Lawrence2020}. The $\phi_i$ are real numbers that have yet to be specified. For every $s_i$, the eigenvalues of $\hat{m}_i (s_i)$ are the elements of $\Omega$. Note that for $s_i=0$, $\hat{m}_i$ is equal to $X$, in analogy to 2-NMQC. 

\begin{figure*}[t]
	\centering
	\includegraphics[width= 0.7 \linewidth]{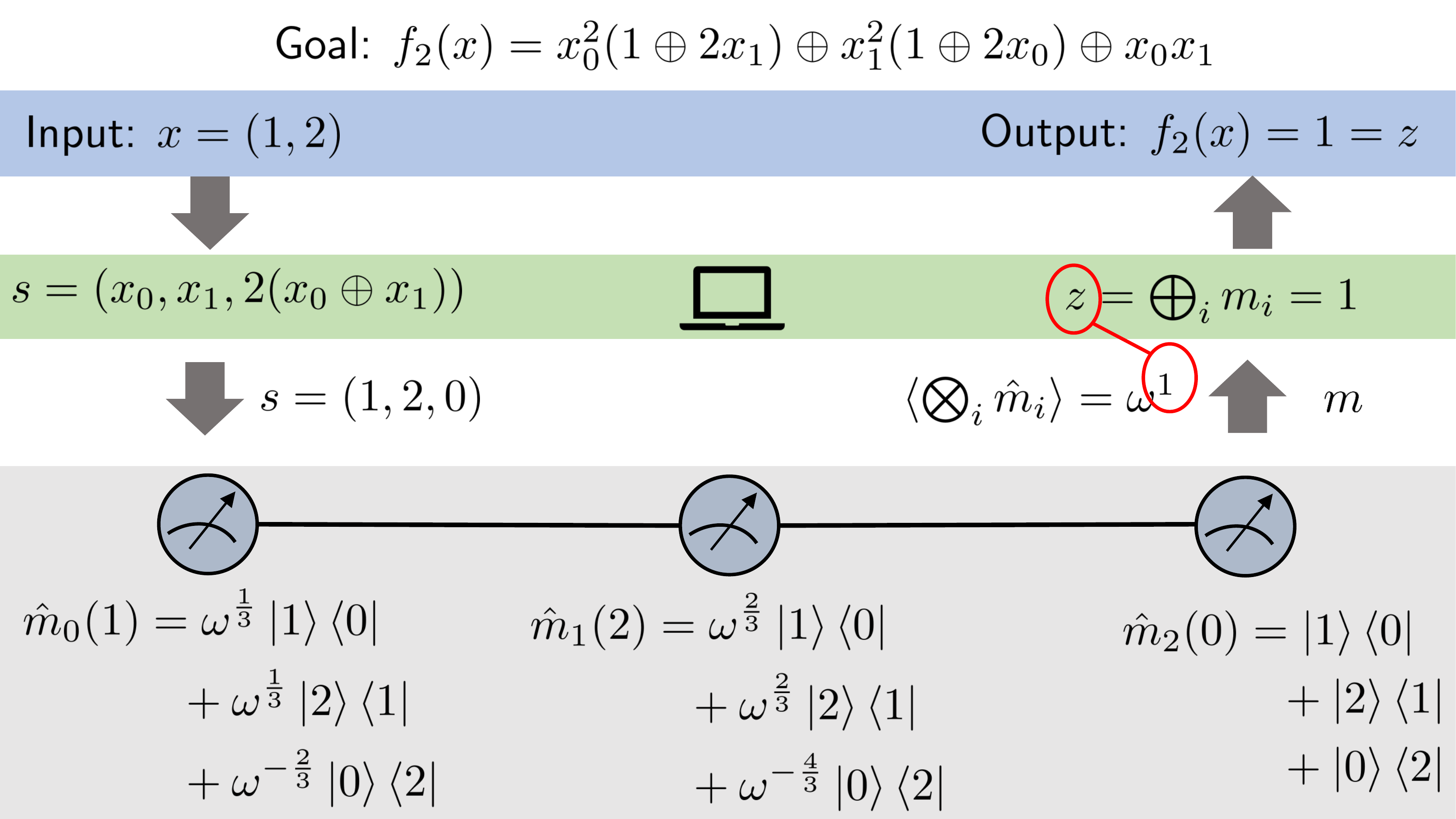}
	\caption{3-NMQC computation of the function $f_2(x)$ for the input $x=(1,2)$. The input is pre-processed using the instructions from Proposition~\ref{prop:exfunc}. Then, the three-qutrit GHZ state (symbolised using graph notation) is measured with respect to the resulting measurement operators $\hat{m}_i(s_i)$ determined by Eqn.~\eqref{eq:mi_func}. The exponent in the measurement result's expectation value $\langle\bigotimes_i \hat{m}_i \rangle$ is equal to $z$. Evaluating $f_2(x=(1,2))$ shows that $z=f(x)$ for this input.}
	\label{fig:3NMQCex}
\end{figure*}

As defined above, the measurement results $M_i$ are mapped to the values $m_i$ via $\omega^{m_i}= M_i$. As a result, if the tensor product $\bigotimes_{i=0}^{l-1} \hat{m}_i$ yields the measurement result $\prod_{i=0}^{l-1} M_i =  \omega^{\bigoplus_{i=0}^{l} m_i} = 1$, it corresponds to $z = 0$ and if it yields $\prod_{i=0}^{l-1} M_i =  \omega^{\bigoplus_{i=0}^{l} m_i} = \omega$, it corresponds to $z=1$, etc. By applying the tensor product $\bigotimes_{i=0}^{l-1} \hat{m}_i$ of the parametrised operators to the qutrit GHZ state defined by Eqn.~\eqref{eq:3GHZ}, we obtain:
\begin{align}
	&\bigotimes_{i=0}^{l-1}\hat{m}_i \GHZketdrei\\
	&=\bigotimes_{i=0}^{l-1}\left(\alpha^{s_i\phi_i}\ket{1}\bra{0} + \alpha^{s_i\phi_i}\ket{2}\bra{1} \right.  \nonumber  \\
	&\phantom{=}  \left.+ \alpha^{-2 s_i \phi_i} \ket{0}\bra{2}\right) \GHZketdrei\\
	&= \frac{1}{\sqrt{3}} \left(\alpha^{\sum_i s_i \phi_i}\ket{0}^{\otimes l}+\alpha^{\sum_i s_i \phi_i}\ket{1}^{\otimes l}\right. \nonumber\\
	&\phantom{=}+ \left. \alpha^{-2 \sum_i s_i \phi_i}\ket{2}^{\otimes l}\right).  \label{eq:GHZmitensorprod} 
\end{align}

For the 3-NMQC procedure to deterministically compute a function $f: \GFtn \rightarrow \GFt$,  Eqn.~\eqref{eq:GHZmitensorprod} must be equal to $\omega^{f(x)} \GHZketdrei$. Therefore, the conditions that have to hold for such a deterministic 3-NMQC scheme are:  
\begin{eqnarray} 
\alpha^{\sum_i s_i \phi_i} &&= \omega^{f(x)+c} \label{eq:ConditionSuccess1} \\
\alpha^{-2 \sum_i s_i \phi_i} &&= \omega^{f(x)+c} , \label{eq:ConditionSuccess2}
\end{eqnarray}
where $c \in \GFt$ is a post-processing trit that can be added after the computation.
Note that the fulfilment of condition~\eqref{eq:ConditionSuccess2} implies the fulfilment of condition~\eqref{eq:ConditionSuccess1}.

On the left hand side of~\eqref{eq:ConditionSuccess1}, in the exponent, there is a \emph{weighted sum of strictly linear functions over $\GFt$ with coefficients $\in \mathbb{R}$.} 
In analogy to~\cite{Hoban2011IOP}, we need to show that the exponent can represent any function $f: \GFtn \rightarrow \GFt$ using at most $l=3^n-1$ terms. The full proof can be found in Appendix~\ref{sec:Appthm}. 


\section{\label{sec:3NMQCSymm} A family of \texorpdfstring{$n$}{n}-variable functions that can be computed with only \texorpdfstring{$n+1$}{n+1} Qutrits}


Given Theorem~\ref{thm:ghzuniversal}, the question arises whether one can find suitable functions that can be computed with a sufficiently small number of qutrits and, thereby, demonstrate that the superiority of quantum correlations in 3-NMQC is experimentally feasible.  
In this section, we show that indeed this is the case and prove that to compute the function:
 \begin{align}
 		\label{eq:func}
		f_n(x)&=
		\bigoplus_{i=0}^{n-1} x_i^2 \oplus \bigoplus_{i=0}^{n-2}\bigoplus_{j=i+1}^{n-1}x_ix_j \oplus 2\cdot \left[\bigoplus_{i=0}^{n-1}\bigoplus_{j \neq i}^{n-1} x_i^2 x_j\right] \nonumber \\& \phantom{{}={}}\oplus \bigoplus_{i=0}^{n-3}\bigoplus_{j=i+1}^{n-2}\bigoplus_{k=j+1}^{n-1}x_ix_jx_k, \quad n\geq 3  ,
\end{align}
one requires no more than $n+1$ qutrits when using the generalised qutrit GHZ state from Eqn.~\eqref{eq:3GHZ} as a resource. Note that $f_2(x)$ is not defined by Eqn.~\eqref{eq:func} due to the sum indices going up to $n-3$. We therefore directly state its definition:
\begin{equation}\label{eq:func2}
	f_2(x) =   x_0^2 \oplus x_1^2 \oplus  x_0x_1 \oplus 2\cdot \left[x_0^2x_1 \oplus x_1^2 x_0 \right] .
\end{equation}

We summarise the result in the following Proposition:

	\begin{proposition}[Computing $f_n(x)$ with 3-NMQC]\label{prop:exfunc}
		In order to compute the function $f_n: \GFtn \rightarrow \GFt$, as defined by Eqn.~\eqref{eq:func}, with 3-NMQC one requires no more than $n+1$ qutrits. To realise the function, one needs the same measurement operators for every qutrit:
		\begin{equation}\label{eq:mi_func}
			\hat{m}_i =\alpha^{s_i}\ket{1}\bra{0}+\alpha^{s_i}\ket{2}\bra{1}+\alpha^{-2 s_i} \ket{0}\bra{2},
		\end{equation}  
with $\alpha=e^{\frac{2\pi i}{9}}$, and the pre-processing $s_i = x_i$ $\forall i: 0 \leq i \leq n-1$, $s_n = 2\cdot \bigoplus_{i=0}^{n-1}x_i$.
	\end{proposition}

This also means that for any function that LHVs cannot deterministically compute, there is a corresponding generalised Bell inequality for $n$ measurement sites and $3^n$ expectation values (given by Eqn.~\eqref{eq:uprealmain}) that quantum correlations maximally violate with the help of the $\GHZketdrei$ state. 

This follows since in~\cite{Frembs2018} it was shown that at most degree 2 functions can be computed deterministically by LHV theories, and the functions above have higher degree.

It can also be seen as a generalised GHZ paradox for all $n$. As a visual example, the computation of $f_2(x)$ with 3-NMQC using the three-qutrit GHZ state for $x=(1,2)$ is illustrated in Fig.~\ref{fig:3NMQCex}.

\begin{table*}
	\def\arraystretch{1.5}%
	\begin{tabular}{lcccccc}
	\toprule
		$\mathbf{x=(x_0, x_1)}$		 	&$(0,0)$ &$(0,1)$ & $(0,2)$ & $(1,1)$ & $(1,2)$ & $(2,2)$  \\ \midrule 
		$\mathbf{f_2(x)}$ 					& 0  & 1  & 1 & 1& 1 & 2    \\  \midrule 
		$\mathbf{x_0\oplus x_1}$ 		& 0 &1 & 2 & 2 & 0 & 1   \\ \midrule 
		$\mathbf{\alpha^{\sum_{i=0}^{1} x_i + 2 \cdot  (x_0\oplus x_1)}}$\hspace{0.8cm}  &$\alpha^0 = \omega^0$\hspace{0.6cm} & $\alpha^{1+2}=\alpha^3 = \omega^1$ \hspace{0.6cm} & $\alpha^{2+1}=\omega^1$ \hspace{0.6cm} & $\alpha^{2+1}=  \omega^1$\hspace{0.6cm} & $\alpha^{3+0}= \omega^1$\hspace{0.6cm} & $\alpha^{4+2}=\alpha^6=\omega^2$   \\\bottomrule
	\end{tabular}
	\caption{\label{tab:funcvaluesn2}
	The table shows the evaluation of the function $f_2(x)$ given by Eqn.~\eqref{eq:func2} for different values of $x=(x_0, x_1) $ with $x_0, x_1 \in \GFt$  as well as the evaluation of the sum $x_0 \oplus x_1$.
 As $f_2(x)$ is invariant under permutation of the inputs, we only need to check for different combinations of $x_0$ and $x_1$.
We also compute the value $\alpha^{\sum_i s_i \phi_i}=\alpha^{\sum_{i=0}^{1} x_i + 2 \cdot  (x_0\oplus x_1)}$, which is the expectation value of the tensor product of the three measurement operators determined by $x_0$ and $x_1$ with regard to the three-qutrit GHZ state.
 We see that, using the pre-processing as defined in Proposition~\ref{prop:exfunc}, the expectation value is always equal to $\omega^{f(x)}$, fulfilling the condition for deterministic 3-NMQC given by Eqn.~\eqref{eq:ConditionSuccess1}. Therefore, 
		for $n=2$, the chosen pre-processing and measurement operators result in the computation of $f_2(x)$. }
\end{table*}


\subsection{Main idea behind the proof of Proposition~\ref{prop:exfunc}} \label{sec:proofsketch2}


We briefly explain the reasoning for the ansatz in Proposition~\ref{prop:exfunc} as it demonstrates the mathematical parallels to 2-NMQC. Its full proof via natural induction can be found in Appendix \ref{sec:Appinduction}. 

For $n=2$, we evaluate $f_2(x)$ for every $x$ and verify that the condition for deterministic 3-NMQC, given by Eqn.~\eqref{eq:ConditionSuccess1}, always holds when using the pre-processing from Proposition~\ref{prop:exfunc}. This is summarised in Table \ref{tab:funcvaluesn2}. 

For $n\geq 3$, the idea behind finding $f_{n}(x)$ and its appropriate measurement operators and pre-processing
follows the pre-processing of its qubit analogue, the pairwise AND function (Eqn.~\eqref{eq:pairwiseand}). 
To compute the pairwise AND function, one chooses the measurement operators $\hat{m}_i(s_i=0) = \sigma_x$ and $\hat{m}_i(s_i=1) = \sigma_y $ for all $i$ and a qubit GHZ state (Eqn.~\eqref{eq:2GHZ}).
The pre-processing is given by (see~\cite{Hoban2011IOP}):
\begin{equation}
	s_i = \begin{cases}
		 x_i & 0 \leq i \leq n-1 \\
		 \bigoplus_{j=0}^{n-1} x_j & i=n
		\end{cases} .
\end{equation}
The resulting tensor product of measurement operators always contain an \emph{even} number of $\sigma_y$ operators, so that any global phase in front of the qubit GHZ state is an element of $\Omega = \{-1, 1\}$. Thus, the expectation value will also always be an element of $\Omega$.

For 3-NMQC, we now choose the qutrit measurement operators given by Eqn.~\eqref{eq:mi_func}.
The pre-processing:
\begin{equation}\label{eq:3prepro}
	s_i = \begin{cases} x_i &  0 \leq i \leq n-1 \\
		  2 \cdot  \left(\bigoplus_j x_j\right) & i = n
		\end{cases}
\end{equation}
results in the following expectation values with respect to the qutrit GHZ state: 
\begin{align} 
	\GHZbradrei \bigotimes_i \hat{m}_i \GHZketdrei &= \alpha^{\sum_{i=0}^{n-1} (x_i)+2\cdot s_n} \label{eq:funcexpprepro1} \\
	&= \alpha^{\sum_{i=0}^{n-1} (x_i)+2\cdot \bigoplus_{i=0}^{n-1}\left(x_i \right)} . \label{eq:funcexpprepro2}
\end{align}
These always consist of \emph{multiples of} $\alpha$ that are \emph{divisible by 3}, i.e. that the expectation value is an element of the set of the third roots of unity $\Omega = \{1, \omega, \omega^{2} \}$ for all $x$.

By evaluating the expectation values given by Eqn. \eqref{eq:funcexpprepro2} for all inputs $x$ for some $n$ and interpolating to find the so-called ``Reed-Muller expansion'' from~\cite{Stankovic2012}, we inferred the expression in Eqn.~\eqref{eq:func} for the function that can be computed with this pre-processing. We prove that this is the correct form for all $n$ in Appendix \ref{sec:Appinduction}. 
It is important to note that, just as in the case of $\GF$ and the pairwise AND function, $f_n(x)$ is invariant under all coordinate permutation.  


\subsection{Generalised qutrit Bell inequality} \label{sec:Prop1Bell}


Let us now discuss the generalised qutrit Bell inequality for the generalised qutrit GHZ paradox induced by the deterministic computation of the function given by Eqn.~\eqref{eq:func}. The quantum bound of this generalised Bell inequality is the maximal $q=1$ due to Proposition~\ref{prop:exfunc} (as explained in Section \ref{sec:3NMQCBellalt}). 
 If we insert $f_2(x)$ from Eqn.~\eqref{eq:func2} and $w(x) =1/3^n$ into Eqn.~\eqref{eq:uprealmain} for the related generalised qutrit Bell inequality, we obtain:
\begin{align}\label{eq:Bellf2}
&\Re{\sum_{x \in \mathbb{F}_2^2} \frac{1}{3^n} \omega^{-f_2(x)} E(z|s=(x_0,x_1,2\cdot(x_0\oplus x_1)))} \nonumber \\
& = \frac{1}{3^n} \Re\Big\{E(z|s=(0,0,0)) \nonumber \\
	&\phantom{=}+\omega^{-1}\sum_{x \in \mathbb{F}_2^2\setminus\{(0,0),(2,2)\}} E(z|s=(x_0,x_1,2\cdot(x_0\oplus x_1))) \nonumber \\
	&\phantom{=}+\omega^{-2} E(z|s=(2,2,2))\Big\} \leq \begin{cases} c = 2/3 \\ q =1 \end{cases} .
\end{align}

 The LHV bound $c$ was found numerically. We also computed $c$ for $n=3$, i.e. for three input trits and $l=4$ qutrits and find that the classical bound of the induced Bell inequality, which we will not explicitly state here, is $c= 1/2$. The bound has decreased by a considerable amount with an increase of the number of qutrits.

It turns out that the inequality \eqref{eq:Bellf2} coincides with one of the Mermin inequalities for 3 qutrits discussed in ~\cite{Lawrence2017}. The LHV bound of the expectation value of the equivalent Mermin inequality is, thereby, in agreement with the LHV bound $c=2/3$ of our normalised Bell inequality given by Eqn.~\eqref{eq:Bellf2}. Therefore, we may say that with $f_n(x)$ in Eqn.~\eqref{eq:func}, for $n>2$, we have found another version of the Mermin operator for 3 qutrits in~\cite{Lawrence2017}.


\section{\label{sec:3NMQCexpscale}Not all functions can be efficiently computed with 3-NMQC}\label{sec:ineff}

One might think that the scalability of the computation of Eqn.~\eqref{eq:func} implies that all functions can be computed efficiently with qutrits and 3-NMQC using the qutrit GHZ state in Eqn.~\eqref{eq:3GHZ}. However, it turns out that there are functions whose computation scales exponentially with the number of input trits. They are analogous to their binary counterparts, in that they differ from a constant function at only one input $x$ (see Section \ref{sec:BG:2NMQC}):

 \begin{proposition}\label{prop:ScalingConstant}
	To compute a function $f: \GFtn \mapsto \GFt$ with a (Hamming) distance 1 to a constant function $f(x)=c$ with $c \in \GFt$ with 3-NMQC and the generalised qutrit GHZ state (defined in \eqref{eq:3GHZ}) as a resource, one needs no fewer than $(3^n-1)/2$ qutrits.
\end{proposition}

\begin{figure}[b]
	\centering
	\includegraphics[width= \linewidth]{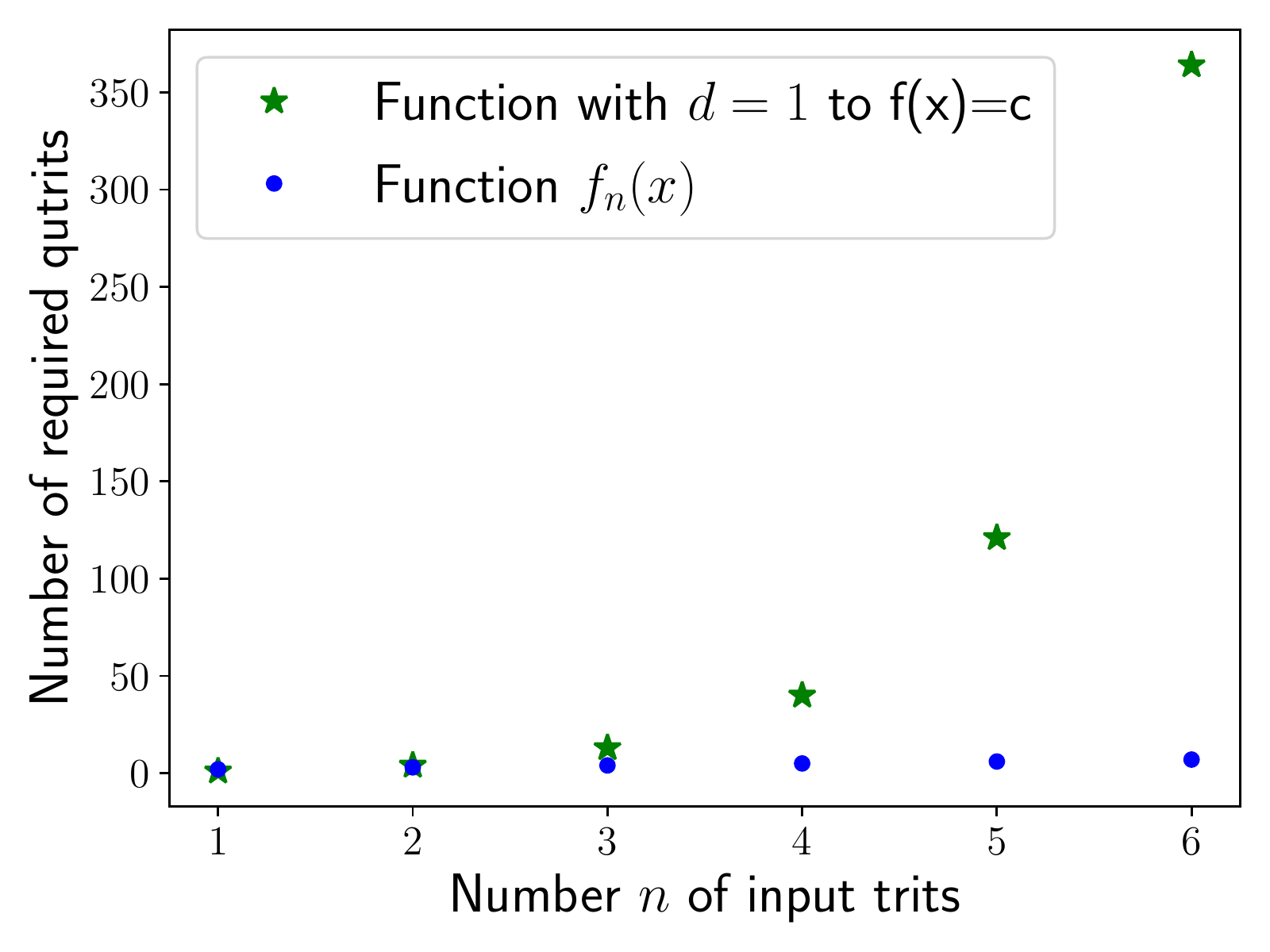}
	\caption{Comparison between the required number of qutrits to compute the functions $f_n(x)$, defined by Eqn. \eqref{eq:func}, and a function with a Hamming distance $d$ of 1 to a constant one with 3-NMQC and the qutrit GHZ state.}
	\label{fig:scaling}
\end{figure}

We prove Proposition \ref{prop:ScalingConstant} in Appendix~\ref{sec:AppPropExp}. 
Just as it was the case for 2-NMQC, Proposition \ref{prop:ScalingConstant} demonstrates that not all functions can be computed efficiently with 3-NMQC and the qutrit GHZ state. 
There is a stark contrast between the required numbers of qutrits for the computation of $f_n(x)$, defined in Eqn.~\eqref{eq:func}, and a function with a Hamming distance of 1 to a constant one when using the qutrit GHZ state. We visualise this difference in scaling for some numbers $n$ of input trits in Fig.~\ref{fig:scaling}.

\section{Possible experimental realisations of 3-NMQC}\label{sec:exp}
 
As we have proven that the generalised qutrit GHZ state is a universal resource for 3-NMQC, it is also the most obvious resource for an experimental implementation of 3-NMQC.
It has also been shown that if a 3-NMQC model can compute a function of a degree greater or equal to 3~\cite{Frembs2018}, it is strongly non-local, following the definition of strong non-locality in~\cite{Abramsky2011}. Thus, the computation of the function given by Eqn. \eqref{eq:func} with qutrits could serve as an experimental demonstration of strong non-locality.

One possible way of realising qutrit systems is to utilise superconducting transmon processors~\cite{Blok2021,Galda2021,Cervera-Lierta2021}. Addressing three energy levels of the transmons, thereby, allows for implementing single- and two-qutrit gates and even the generation of a three-qutrit GHZ state has been demonstrated~\cite{Cervera-Lierta2021}.
In a similar way, quantum processors with trapped ions benefit from the multi-level structure of the ions, which has led to the implementation of single-qutrit gates~\cite{Ringbauer2021}.

Another promising system to generate and measure qutrit GHZ states is the photonic system~\cite{Lee2005,Erhard2018,Paesani2021,Bell2021}, which has already been used to implement 2-NMQC ~\cite{Demirel2021}. Naturally, photonic degrees of freedom such as path, orbital angular momentum (OAM), or photon number, can be of arbitrary dimensions. Employing OAM, the authors of~\cite{Erhard2018} have generated a three-qutrit GHZ state and shown that it would be suited to violate the Bell/Mermin inequality given in~\cite{Lawrence2017}. It would therefore realise the computation of the function $f_2(x)$ [see Eqn.~\eqref{eq:func2}]. The necessary single-qutrit measurement operations have been discussed in \cite{Lee2005} and \cite{Babazadeh2017} for path and OAM degrees of freedom, respectively, and could be implemented in future experiments. One could further generalise these setups to realise the computation of the function $f_n(x)$ defined by Eqn. \eqref{eq:func} for all $n$.

\section{Conclusion and Outlook}\label{sec:concl}

We have derived a series of novel findings on non-adaptive measurement-based quantum computing (NMQC) with qutrits and ternary logic (3-NMQC).

We showed that with a qutrit GHZ state as a resource, a restricted ternary pre-processor can be boosted to compute any ternary function in 3-NMQC.
The associated generalised Bell inequalities are maximally violated by qutrit GHZ states.
We showed that the minimum size of the qutrit GHZ state that is required to compute a function that differs from a constant function at only one input value scales exponentially with $n$.

We also presented a family of $n$-variate functions which require only $n+1$ qutrits using the qutrit GHZ state for NMQC.
Interestingly, the probabilistic NMQC game induced by the equivalent function in the binary case, the pairwise AND function, is related to a Svetlichny inequality~\cite{Hoban2011IOP}. There, the ratio of the quantum to the classical bound q/c increases exponentially with the number of qubits.
We conjecture that our function may also be related to a generalised Svetlichny inequality for qutrits.

It is an interesting open question whether the generalised Bell inequalities induced by the probabilistic 3-NMQC game form a \emph{complete} set of Bell inequalities.
 The binary NMQC game can describe the entire, \emph{complete} set of qubit $(n,2,2)$ Bell inequalities \cite{Hoban2011IOP} which are maximally and optimally violated by the generalised qubit GHZ state. Due to the similarities between binary and 3-NMQC, the 3-NMQC game might also describe an entire \emph{complete} set of generalised qutrit Bell inequalities, which are also \emph{optimally} and \emph{maximally} violated by the generalised qutrit GHZ state.

 A natural extension of our work is general $p$-NMQC, where the goal is to compute a function $f: \GFpn \rightarrow \GFp$. It is interesting to see whether our proof can be extended to $p>3$ and whether quantum correlations can indeed elevate any pre-processor to classical universality. Furthermore, it is an interesting question whether the parallel between binary and ternary NMQC with regard to the functions that cannot be efficiently computed remains for $p$-NMQC in general. \\


\begin{acknowledgments}
We thank David Canning for comments on the manuscript.
We acknowledge support from the Carl Zeiss Foundation, the Centre for Integrated Quantum Science and Technology (IQ$^\text{ST}$), the German Research Foundation (DFG), the Federal Ministry of Education and Research (BMBF, projects SiSiQ and PhotonQ), and the Federal Ministry for Economic Affairs and Energy (BMWi, project PlanQK), the Competence Center Quantum Computing Baden-Württemberg (funded by the Ministerium für Wirtschaft, Arbeit und Tourismus Baden-Württemberg, project QORA). MJH now works for Quantinuum, but did not contribute to this work while at his current affiliation.
\end{acknowledgments}


%
%
%

%


%
%
%

 \appendix
\newpage
\begin{widetext}
	
%
\section{Derivation of the alternative generalised Bell inequality (\ref{eq:uprealmain})}\label{sec:AppBell}
%

We start out with the expectation values:
\begin{equation}
	\label{eq:EMapp}
	E^{M}(z|s) = \sum_{k=0}^{2} \omega^{kM} p(z=k|s).
\end{equation} 

Now, we note that, multiplying $\omega^{-f(x)}$ with $E^{1}(z|s) \equiv E(z|s)$ leads to:
\begin{equation} \label{eq:omfE}
		\omega^{-f(x)} E(z|s) = p(z=f(x)|s) + \omega^1 p(z=f(x)\oplus 1|s) +\omega^2 p(z=f(x)\oplus 2|s).
\end{equation}

If we now sum up Eqn.~\eqref{eq:omfE} over all $x$, weighted by $w(x)$ and discard the imaginary part, we arrive at:
	\begin{align} 
		&\Re{\sum_x w(x) \omega^{-f(x)} E(z|s)} \label{eq:omfEsum1}  \\
		&= \Re{\sum_x w(x) \left(p(z=f(x)|s) + \omega^1 p(z=f(x)\oplus 1|s) +\omega^2 p(z=f(x)\oplus 2|s)\right)} \label{eq:omfEsum2} \\
		&= \sum_x w(x)  p(z=f(x)|s) -\frac{1}{2} \left(p(z=f(x) \oplus 1|s) +p(z=f(x)\oplus 2|s)\right) \label{eq:omfEsum3}\\
		& = \ps -\frac{1}{2} \sum_x w(x) p(z\neq f(x)|s)  \label{eq:omfEsum4} \\
		& = \ps -\frac{1}{2}(1-\ps)  \label{eq:omfEsum5} \\
		& = \frac{3}{2} \ps -\frac{1}{2}.  \label{eq:omfEsum6}
	\end{align}
 
In Eqn.~\eqref{eq:omfEsum4} we introduced the probability $ p(z\neq f(x)|s) = \left(p(z=f(x) \oplus 1|s) +p(z=f(x)\oplus 2|s)\right)$ that $z$ is \emph{not} equal to $f(x)$ given that the measurement settings are determined by $s$. In Eqn.~\eqref{eq:omfEsum3} we insert the definition of $\ps$ (see Eqn.~\eqref{eq:psdef3}). We now redefine $u_3$ as:
\begin{equation}\label{eq:upreal}
	u_3 =  \sum_{x} w(x) \omega^{-f(x)}E(z|s)     ,
\end{equation}
and insert it into Eqn.~\eqref{eq:omfEsum1}:
\begin{equation} \label{eq:psBell3}
	\ps = \frac{1}{3} (1+ 2 \Re{u_3}), \qquad \Re{u_3} \leq   \begin{cases} c \\ q \end{cases} .
\end{equation}
\section{Proof of Theorem \ref{thm:ghzuniversal}} \label{sec:Appthm}

As outlined in section \ref{sec:proofsketch}, the idea behind the proof of Theorem~\ref{thm:ghzuniversal} is to show that the exponent on the l.h.s. of the condition for deterministic 3-NMQC: 
\begin{equation}
\alpha^{\sum_i s_i \phi_i}  =  \omega^{f(x)+c} \, \, \forall x \in \GFtn, \label{eq:ConditionSuccess1app}
\end{equation}
which is a sum in $\mathbb{R}$ of strictly linear functions, can represent any function for $l \leq 3^n-1$.  Here, we complete the proof. \newline
The idea of the proof is relatively simple. Observe that if we write out Eqn.~\eqref{eq:ConditionSuccess1app} for every $x$, we have a system of $3^n$ equations that have to hold true. On both sides there is an $\omega$ in the base ($\alpha=\omega^{\frac{1}{3}}$) and on the left hand side there is a sum over the reals of strictly linear functions defined and evaluated in $\GFt$ in the exponent. All $3^n$ equations are true if the exponents on both sides are always equal. This would mean that (absorbing the factor $1/3$ into $\phi_{i}$, i.e., $\phi_{i}/3 \mapsto \phi_{i}$):
\begin{equation}
	 \sum_i s_i \phi_i   =   f(x) \oplus c  \, \, \forall x \in \GFtn. \label{eq:ConditionSuccess1exp}
\end{equation}

We can now interpret the set of these $3^n$ equations as one single condition for the \emph{function vector} $\vec{f} = (f(x=(0,0,\ldots,0)),f(x=(0,0,\ldots,1)), \ldots , f(x=(2,2,\ldots,2)))^T$ of $f: \GFtn \rightarrow \GFt$, which is the sorted image of $f$. On the left hand side, we then have a sum in the reals over the function vectors of the strictly linear functions:
\begin{equation}
	\sum_{i} \phi_i \vec{f}_{v^{(i)}}   =   \vec{f} \oplus (c,c,...,c)^T, \label{eq:ConditionSuccess1func}
\end{equation}
whereby we write strictly linear functions as a scalar product:
\begin{equation}
	f(x) = v \cdot x, \quad v \in \GFtn
\end{equation}
and index its function vector with the vector $v \in \GFtn$ that defines it.

 Thus, what we need to show is that the function vectors of the (strictly) linear functions in $\GFp$, augmented by the function vector of the constant function $f(x)=c \; \forall x, \, c \in \{1,2\}$, (realised by the post-processing trit) can represent any vector in $\mathbb{R}^{3^n}$. There are $3^n-1$ non-trivial vectors $v \in \GFtn$ and thus $3^n-1$ non-trivial (strictly) linear functions. If we augment this set by the function vector $\vec{f}_c$ of the constant function, we arrive at a set of $3^n$ vectors. As we want to represent any vector in a $3^n$-dimensional vector space over $\mathbb{R}$, it only remains to show that the set of function vectors of strictly linear functions, augmented by the function vector of the constant one, is linearly independent over $\mathbb{R}$. \newline
 In Lemma \ref{lem:lem} we show exactly that. Then, we directly conclude that Theorem~\ref{thm:ghzuniversal} holds due to the reasoning outlined above.

%
\subsection{A useful Lemma}
%

We begin with lemma \ref{lem:lem}:
 
	\begin{lemma} \label{lem:lem}
		Let $\vec{f}_v$ denote the function vector of a non-trivial strictly linear function $f: \GFtn \rightarrow \GFt:: x \mapsto \bigoplus_{i=0}^{n-1} v_i x_i, \quad v \neq \vec{0}$, i.e.:
		$\vec{f}_v = (f((0,0,...,0)^T),f((0,0,...,1)^T)), f((0,0,...,2)^T)^T, ...,f((2,2,...,2)^T)$. The vectors in the set $\{\vec{f}_v : v \in \GFtn\}$, augmented by the function vector $\vec{f}_c = (1,1,1,1,..,1)^T$ of the constant function $f(x)=1$, are linearly independent over $\mathbb{R}$.
	\end{lemma} 
 
\paragraph*{Proof of lemma \ref{lem:lem}} 
In the following, we abbreviate $\bigoplus_{i=0}^{n-1} v_i x_i$ by $v \cdot x$.\newline
To prove lemma \ref{lem:lem}, we must show that the following equation:
\begin{equation} \label{eq:LinIndepCond}
\sum_{v \in \GFtn, v \neq 0} \lambda_v \vec{f_v}  + \lambda_c \vec{f}_\mathrm{const}= 0, \quad \lambda_v \in \mathbb{R} \forall \lambda_v, \, \lambda_c \in \mathbb{R}
\end{equation}
leads to $\lambda_v = 0 \, \forall v \land \lambda_c = 0$. In the sum subscript, we wrote $\vec{v} \neq \vec{0}$ as $v \neq 0$. \newline
First, all strictly linear functions are 0 at $x=\vec{0}$. Therefore, the first line of the linear system of equations Eqn.~\eqref{eq:LinIndepCond} is:
\begin{equation}
0+0+...+0+\lambda_c = 0 \implies \lambda_c = 0.
\end{equation}

We can split up all elements of $\GFtn$ except the zero vector into sets of $(3^n-1)/2$ sets dubbed $X$ and $2X$ in which no element is the additive inverse of another one. Then, the following holds true:
\begin{equation}
\forall x \in X \exists y \in 2X: y = 2x.
\end{equation}
(We are still talking about element-wise modulo 3 addition and scalar multiplication.) 

If a vector $v \in \GFtn \setminus {\vec{0}}$ fulfils the condition $v \cdot x = 1$ for some $x \in X$, then one can find the (unique) additive inverse $y \in 2X$ with $y = 2x$ and it follows that $v  \cdot y = v \cdot 2x = 2$. Let us get back to the linear system of equations Eqn.~\eqref{eq:LinIndepCond}. For an $x \in X$ we can split the sum up into three terms:
	\begin{equation} \label{eq:SplitLambdaSumX}
	\sum_{v \in \GFtn \setminus \{0\}} \lambda_v  (v \cdot x)   = 
	\sum_{v \in \GFtn \setminus \{0\}, v \cdot x = 0} \lambda_v \cdot  0  +  
	\sum_{v \in \GFtn \setminus \{0\}, v \cdot x = 1} \lambda_v  \cdot 1  +  
	\sum_{v \in \GFtn \setminus \{0\}, v \cdot x = 2} \lambda_v  \cdot 2 = 0.
	\end{equation} 
We can do the same with the corresponding inverse $y = 2x$ from the set $2X$:
	\begin{equation} \label{eq:SplitLambdaSum2X}
	\sum_{v \in \GFtn \setminus \{0\}} \lambda_v  (v \cdot y)    = 
	\sum_{v \in \GFtn \setminus \{0\}, v \cdot y = 0} \lambda_v \cdot  0  +  
	\sum_{v \in \GFtn \setminus \{0\}, v \cdot y = 1} \lambda_v  \cdot 1 + 
	\sum_{v \in \GFtn \setminus \{0\}, v \cdot y = 2} \lambda_v  \cdot 2  = 0. 
	\end{equation} 
Obviously, if $v \cdot x = 0$ then $v \cdot 2x = 2 (v \cdot x) = 0$. (This is true for addition and multiplication modulo 3 as well.) So the coefficients in the first term on the right hand sides of Eqn.~\eqref{eq:SplitLambdaSumX} and Eqn.~\eqref{eq:SplitLambdaSum2X} are the same: 
\begin{equation}
\{\lambda_v: v \cdot x = 0\} = \{\lambda_w: w \cdot 2x = 0\}.
\end{equation}
We are left with the terms where $v \cdot x$ equals 1 and 2. If $(2x) \cdot v =1$, then $x \cdot v = 2$ and if $(2x) \cdot v =2$, then $x \cdot v = 1$. So one can rewrite equations~\eqref{eq:SplitLambdaSumX} and~\eqref{eq:SplitLambdaSum2X} as:
	\begin{equation}
\sum_{v \in \GFtn \setminus \{0\}} \lambda_v  (v \cdot x)  
= 	\sum_{v \in \GFtn \setminus \{0\}, v \cdot x = 1} \lambda_v \cdot  1  + 
	\sum_{v \in \GFtn \setminus \{0\}, v \cdot x = 2} \lambda_v  \cdot 2  = 0
	\end{equation}
and
	\begin{align} 
	\begin{split}
	\sum_{v \in \GFtn \setminus \{0\}} \lambda_v  (v \cdot y)   
	&= \sum_{v \in \GFtn \setminus \{0\}, v \cdot y = 1} \lambda_v  \cdot 1 +  
	\sum_{v \in \GFtn \setminus \{0\}, v \cdot y = 2} \lambda_v  \cdot 2 \\
	&=\sum_{v \in \GFtn \setminus \{0\}, v \cdot (2x) = 1} \lambda_v  \cdot 1 +  
	\sum_{v \in \GFtn \setminus \{0\}, v \cdot (2x) = 2} \lambda_v  \cdot 2 \\
	&=\sum_{v \in \GFtn \setminus \{0\}, v \cdot x = 2} \lambda_v  \cdot 1 + 
	\sum_{v \in \GFtn \setminus \{0\}, v \cdot x = 1} \lambda_v  \cdot 2 
	= 0.
	\end{split}
	\end{align}

We end up with the following two conditions for all $x \in X$:
	\begin{align} 
	\sum_{v \in \GFtn \setminus \{0\}, v \cdot x = 2} \lambda_v  \cdot 1 + \sum_{v \in \GFtn \setminus \{0\}, v \cdot x = 1} \lambda_v  \cdot 2 &= 0  \\
	\sum_{v \in \GFtn \setminus \{0\}, v \cdot x = 1} \lambda_v  \cdot 1 + \sum_{v \in \GFtn \setminus \{0\}, v \cdot x = 2} \lambda_v  \cdot 2 &= 0
	\end{align} 
 
They have to hold for $\lambda_v \in \mathbb{R} \forall v$ and addition and multiplication in $\mathbb{R}$ (\textbf{not} $\GFt$!). This is only possible if:
\begin{equation} \label{eq:LambdaSum1and2isZero}
\sum_{\substack{v \in \GFtn \setminus \{0\} \\ v \cdot x = 1}}\lambda_v =  \sum_{\substack{v \in \GFtn \setminus \{0\} \\ v \cdot x = 2}} \lambda_v = 0.
\end{equation}
Therefore:
\begin{equation}  
\sum_{\substack{v \in \GFtn \setminus \{0\} \\ v \cdot x = 0}}\lambda_v 
	=	\sum_{\substack{v \in \GFtn \setminus \{0\}}}\lambda_v - \sum_{\substack{v \in \GFtn \setminus \{0\} \\ v \cdot x = 1}}\lambda_v -\sum_{\substack{v \in \GFtn \setminus \{0\} \\ v \cdot x = 2}}\lambda_v  
= \sum_{\substack{v \in \GFtn \setminus \{0\}}}\lambda_v \forall x.
	 \label{eq:LambdaSum0IsTheSame}
\end{equation}
 
So the conditional sum over all $v$ that fulfil $v \cdot x = 0$ for some $x \in \GFtn \setminus \{0\}$ is the same for every $x$. \newline 
For a given $v \in \GFtn \setminus \{0\}$, how many $x$ are there that fulfil $v \cdot x  = 1$? \newline
Every $x$ has $n$ elements. Without loss of generality, we can fix one index $k$ for which $v_k$ is not zero (There exists at least one such index if $v$ is non-zero.) and keep the corresponding $x_k$ fixed. We can then freely choose all other $n-1$ elements of $x$ and then have to pick the left over $x_k$ such that $v \cdot x = 1$.  The  $n-1$ elements of $x$ that can be freely chosen lead to $3^{n-1}$ $x \in F_3^n \setminus \{0\}$ that fulfil $x \cdot v =1$ for a given $v \in F_3^n \setminus \{0\}$. The same is true for the condition $v \cdot x = 2$. Keeping this in mind, we sum one of the terms in Eqn.~\eqref{eq:LambdaSum1and2isZero} over all $x \in F_3^n \setminus\{0\}$:
 \begin{align}
	\begin{split}
	\sum_{x \in \GFtn \setminus \{0\}}  \left(\sum_{\substack{v \in \GFtn \setminus \{0\} \\ v \cdot x = 1}}\lambda_v \right) &= \sum_{x \in \GFtn \setminus \{0\}}  0 
	=\sum_{x \in \GFtn \setminus \{0\}}   \left(\sum_{\substack{v \in \GFtn \setminus \{0\}}} \delta_{v \cdot x =1}\lambda_v \right) \\
	&=  \sum_{v \in \GFtn \setminus \{0\}}   \left(\sum_{\substack{x \in \GFtn \setminus \{0\}}} \delta_{v \cdot x =1}\lambda_v \right)
	=  \sum_{v \in \GFtn \setminus \{0\}}   3^{n-1}\lambda_v  
	=3^{n-1} \sum_{v \in \GFtn \setminus \{0\}}    \lambda_v,
	\end{split}
\end{align}
where $ \delta_{v \cdot x =1}$ is 1 if $v \cdot x = 1$ and 0 otherwise.

Remembering Eqn.~\eqref{eq:LambdaSum0IsTheSame} leads to the conclusion:
\begin{equation} \label{eq:LambdaSum0IsZero} 
\sum_{ v \in F_3^n \setminus \{0\},  v \cdot x = 0}\lambda_v = 0 \quad \forall x.
\end{equation}

Now, we take an arbitrary $x \in \GFtn \setminus \{0\}$ and split it up into $a = (x_0,x_1,...,x_k)$ and $b = (x_{k+1},x_1,...,x_{n-1})$ for some $k: 1 \leq k \leq n-2$. Eqn.~\eqref{eq:LambdaSum0IsZero} leads us to:
\begin{align} \label{eq:EquationToShowSplitSumZero}
\begin{split}
\sum_{\substack{v \in \GFtn \setminus \{0\} \\ v \cdot x = 0}}\lambda_v   &= 
\sum_{\substack{v \in \GFtn \setminus \{0\} \\ v \cdot a = 1 \\ v \cdot b = 2}}\lambda_v +\sum_{\substack{v \in \GFtn \setminus \{0\} \\ v \cdot a = 2 \\ v \cdot b = 1}}\lambda_v + \sum_{\substack{v \in \GFtn \setminus \{0\} \\ v \cdot a = 0 \\ v \cdot b = 0}}\lambda_v \\
&=\sum_{\substack{v \in \GFtn \setminus \{0\} \\ v \cdot a = 1 }}\lambda_v  - 	\sum_{\substack{v \in \GFtn \setminus \{0\} \\ v \cdot a = 1 \\ v \cdot b = 1 }}\lambda_v - 	\sum_{\substack{v \in \GFtn \setminus \{0\} \\ v \cdot a = 1 \\ v \cdot b = 0 }}\lambda_v 
+\sum_{\substack{v \in \GFtn \setminus \{0\} \\ v \cdot a = 2 \\ v \cdot b = 1}}\lambda_v + \sum_{\substack{v \in \GFtn \setminus \{0\} \\ v \cdot a = 0 \\ v \cdot b = 0}}\lambda_v 
\end{split}
\end{align}

The sum $\sum_{\substack{v \in \GFtn \setminus \{0\} \\ v \cdot a = 1 }}\lambda_v$ on the l.h.s. in the second line of Eqn.~\eqref{eq:EquationToShowSplitSumZero} is zero due to Eqn.~\eqref{eq:LambdaSum1and2isZero}. We continue by splitting up other sums analogously to how we split up $\sum_{\substack{v \in \GFtn \setminus \{0\} \\ v \cdot a = 1 \\ v \cdot b = 2}}\lambda_v$: 
	\begin{align}
	\sum_{\substack{v \in \GFtn \setminus \{0\} \\ v \cdot x = 0}}\lambda_v    
	&=0   - 	\sum_{\substack{v \in \GFtn \setminus \{0\} \\ v \cdot a = 1 \\ v \cdot b = 1 }}\lambda_v - 	\sum_{\substack{v \in \GFtn \setminus \{0\} \\ v \cdot a = 1 \\ v \cdot b = 0 }}\lambda_v +\sum_{\substack{v \in \GFtn \setminus \{0\} \\ v \cdot b = 1 }}\lambda_v - \sum_{\substack{v \in \GFtn \setminus \{0\} \\ v \cdot a = 1 \\ v \cdot b = 1 }}\lambda_v -\sum_{\substack{v \in \GFtn \setminus \{0\} \\ v \cdot a = 0 \\ v \cdot b = 1 }}\lambda_v \\
	&\phantom{=}+ \sum_{\substack{v \in \GFtn \setminus \{0\} \\ v \cdot a = 0}}\lambda_v- \sum_{\substack{v \in \GFtn \setminus \{0\} \\ v \cdot a = 0 \\ v \cdot b = 1 }}\lambda_v - \sum_{\substack{v \in \GFtn \setminus \{0\} \\ v \cdot a = 0 \\ v \cdot b = 2 }}\lambda_v \\
	&= - 	\sum_{\substack{v \in \GFtn \setminus \{0\} \\ v \cdot a = 1 \\ v \cdot b = 1 }}\lambda_v - 	\sum_{\substack{v \in \GFtn \setminus \{0\} \\ v \cdot a = 1 \\ v \cdot b = 0 }}\lambda_v - \sum_{\substack{v \in \GFtn \setminus \{0\} \\ v \cdot a = 1 \\ v \cdot b = 1 }}\lambda_v -\sum_{\substack{v \in \GFtn \setminus \{0\} \\ v \cdot a = 0 \\ v \cdot b = 1 }}\lambda_v - \sum_{\substack{v \in \GFtn \setminus \{0\} \\ v \cdot a = 0 \\ v \cdot b = 1 }}\lambda_v - \sum_{\substack{v \in \GFtn \setminus \{0\} \\ v \cdot a = 0 \\ v \cdot b = 2 }}\lambda_v \\
	&=-2 \sum_{\substack{v \in \GFtn \setminus \{0\} \\ v \cdot a = 1 \\ v \cdot b = 1 }}\lambda_v -2 \sum_{\substack{v \in \GFtn \setminus \{0\} \\ v \cdot a = 0 \\ v \cdot b = 1 }}\lambda_v - \sum_{\substack{v \in \GFtn \setminus \{0\} \\ v \cdot a = 1 \\ v \cdot b = 0 }}\lambda_v - \sum_{\substack{v \in \GFtn \setminus \{0\} \\ v \cdot a = 0 \\ v \cdot b = 2}}\lambda_v  \\
	&=-2\left(\sum_{\substack{v \in \GFtn \setminus \{0\} \\ v \cdot b = 1 }}\lambda_v - \sum_{\substack{v \in \GFtn \setminus \{0\} \\ v \cdot a = 2 \\ v \cdot b = 1 }}\lambda_v  \right)   -\left(\sum_{\substack{v \in \GFtn \setminus \{0\} \\ v \cdot (a+2b)= 1 }}\lambda_v -\sum_{\substack{v \in \GFtn \setminus \{0\} \\ v \cdot a = 2 \\ v \cdot b = 1 }}\lambda_v  \right)  \\
	&=-2\left(0 - \sum_{\substack{v \in \GFtn \setminus \{0\} \\ v \cdot a = 2 \\ v \cdot b = 1 }}\lambda_v  \right)   -\left(0 -\sum_{\substack{v \in \GFtn \setminus \{0\} \\ v \cdot a = 2 \\ v \cdot b = 1 }}\lambda_v  \right) \\
	&=3 \sum_{\substack{v \in \GFtn \setminus \{0\} \\ v \cdot a = 2 \\ v \cdot b = 1 }}\lambda_v  = 0.
	\end{align}
(All sums $\sum_{\substack{v \in \GFtn \setminus \{0\} \\ v \cdot y = m}}\lambda_v$ for some $y \in \GFtn$ and some $m \in \GFt$ vanish due to Eqn.~\eqref{eq:LambdaSum1and2isZero}.)
In the same manner, one can show that for any $m,n \in \GFt$ and any $a, b$ defined as above:
\begin{equation} \label{eq:LambdaSplitSumZero}
\sum_{\substack{v \in \GFtn \setminus \{0\} \\ v \cdot a = m \\ v \cdot b = n }}\lambda_v = 0.
\end{equation}

We can now split up the vector $a = (x_0,....,x_k)$ into $a'=(x_0,...,x_l)$ and $\tilde{a} = (x_{l+1},...,x_l)$ for some arbitrary $l$ with $1 \leq l \leq k-1$ and rewrite the sum in Eqn.~\eqref{eq:LambdaSplitSumZero} as:
\begin{equation}  
\sum_{\substack{v \in \GFtn \setminus \{0\} \\ v \cdot a = m \\ v \cdot b = n }}\lambda_v =   
\sum_{\substack{v \in \GFtn \setminus \{0\} \\ v \cdot a' = m \\ v \cdot \tilde{a} = 0 \\v \cdot b = n }}\lambda_v + 	\sum_{\substack{v \in \GFtn \setminus \{0\} \\ v \cdot a' = m+1 \\ v \cdot \tilde{a} = 2 \\v \cdot b = n }}\lambda_v+ 	\sum_{\substack{v \in \GFtn \setminus \{0\} \\ v \cdot a' = m+2 \\ v \cdot \tilde{a}= 1 \\v \cdot b = n }}\lambda_v.
\end{equation}

Following the same procedure as from the second line of Eqn.~\eqref{eq:EquationToShowSplitSumZero} onwards, one arrives at:
\begin{equation} \label{eq:LambdaSplitTwiceSumZero}
\sum_{\substack{v \in \GFtn \setminus \{0\} \\ v \cdot a' = m' \\v \cdot \tilde{a} = \tilde{m}   \\ v \cdot b = n}}\lambda_v = 0,
\end{equation}
for all $m',\tilde{m},n \in \GFt$.

Continuing in the same manner, i.e.  splitting $x$ up into increasingly smaller, non-overlapping segments, we end up with the following:
\begin{equation} \label{eq:LambdaSplitManyTimesSumZero}
\sum_{\substack{v \in \GFtn \setminus \{0\} \\ v_0\cdot  x_0 = y_0 \\v_1 \cdot x_1 = y_1  \\...\\ v_{n-1} \cdot x_{n-1} =y_{n-1}}}\lambda_v = 0
\end{equation}
for all $x,y,v \in \GFtn$. Evidently, Eqn.~\eqref{eq:LambdaSplitManyTimesSumZero} can only be true for all $v$ and all $x$ if $\lambda_v = 0 \quad \forall v \in \GFtn \setminus \{0\}$. This completes the proof of lemma \ref{lem:lem}. \qed

With this, we have shown that any function $f: \GFtn \mapsto \GFt$ can be written as a weighted sum of (strictly) linear functions over $\GFt$ with coefficients in $\mathbb{R}$. Therefore, every function can be realised in the exponent on the l.h.s of Eqn.~\eqref{eq:ConditionSuccess1app}. Each strictly linear function stems from a measurement setting $s_i$ and the real coefficient is the angle $\phi_i$ (see Eqn.~\eqref{eq:ConditionSuccess1app}.) (The constant function can be realised by the post-processing trit in Eqn.~\eqref{eq:ConditionSuccess1app}.) There are $3^n-1$ non-trivial strictly linear functions left over, hence the maximal number of $3^n-1$ qutrits. This finalises the proof of Theorem~\ref{thm:ghzuniversal}. \qed

One might object to allowing arbitrary real numbers as angles. We can consider rational approximations of the real numbers, but the result still holds.

%
\section{Proof of Proposition \ref{prop:exfunc}} \label{sec:Appinduction}
%
 
 We prove Proposition~\ref{prop:exfunc} by natural induction.  \newline
 Let us begin with the induction hypothesis. We would like to show that:
 
 \begin{equation} \label{eq:indhyp}
 \alpha^{\sum_{i=0}^{n-1} (x_i)+2\cdot \bigoplus_{i=0}^{n-1}\left(x_i \right)} = \omega^{f_n(x)+c} \quad \forall x,n.
 \end{equation}
 As $f_n(x)$ is invariant under permutation, one only needs to compute the function values for different combinations of $n$ integers taking on values in $\GFt$. For example, for $n=2$, one needs to compute both the left and the right hand side of Eqn.~\eqref{eq:indhyp}, but only for $x=(0,0)$, $x=(0,1)$, $x=(0,2)$, $x=(1,1)$, $x = (1,2)$ and $x = (2,2)$. \newline We choose $n=3$ for the first step of the induction due to the fact that $f_2(x)$ has a separate definition. The reader can easily check for themself that Eqn.~\eqref{eq:indhyp} is fulfilled for $n=2$. \newline
 
 \begin{enumerate}
 	\item $n=3$. \\
 	The l.h.s. of Eqn.~\eqref{eq:indhyp} is $\alpha^0$ for $x=(0,0,0)$, hence $c$ must be zero. For $x=(0,0,1)$, the l.h.s. yields $\alpha^{1+2}=\alpha^3=\omega^1$. Now, we must compute the function value of $f_3$ at $x=(0,0,1)$. All terms including $x_1$ and $x_2$ can be ignored, because these two variables are zero. This means that $f_3(x=(0,0,1))$ is given by:
 	\begin{equation}
 	f_3(x) = x_0^2 =1,
 	\end{equation}
 	which means that Eqn.~\eqref{eq:indhyp} is fulfilled for $n=3$ and $x=(0,0,1)$. All other function values and terms on the left hand side of Eqn.~\eqref{eq:indhyp} are listed in table \ref{tab:funcvaluesn3}. It shows that the induction hypothesis Eqn.~\eqref{eq:indhyp} is correct for $n=3$. 
 \end{enumerate}
 
 \begin{table*}
\bgroup 
\def\arraystretch{1.5}
 	\begin{tabular}{lccccccccc}
	\toprule
 		$\mathbf{x=(x_0, x_1, x_2)}$ &$\hspace{0.2cm}(0,0,1)\hspace{0.2cm}$ & $\hspace{0.2cm}(0,0,2)\hspace{0.2cm}$ &$\hspace{0.2cm}(0,1,1)\hspace{0.2cm}$ & $\hspace{0.2cm}(0,1,2)\hspace{0.2cm}$ & $\hspace{0.2cm}(0,2,2)\hspace{0.2cm}$ & $\hspace{0.2cm}(1,1,1)\hspace{0.2cm}$ & $\hspace{0.2cm}(1,1,2)\hspace{0.2cm}$ & $\hspace{0.2cm}(1,2,2)\hspace{0.2cm}$ &$\hspace{0.2cm}(2,2,2)\hspace{0.2cm}$ \\ \midrule
 		$\mathbf{f_3(x)}$ & 1 & 1 & 1  & 1 & 2& 1 & 2 & 2 &2  \\ \midrule
 		$\mathbf{\bigoplus_{i=0}^{2} x_i}$ & 1& 2 &2 & 0 & 1 & 0 & 1 & 2 & 0  \\ \midrule
 		$\mathbf{\alpha^{\sum_{i=0}^{2} x_i + 2 \cdot \bigoplus_{i=0}^{2} x_i}}$ & $\alpha^3=\omega^1$ &$\omega^1$ &$\omega^1$ & $\omega^1$ & $\alpha^6 = \omega^2$ & $\omega^1$ & $\omega^2$ & $ \omega^2$ &$\omega^2$
		\\\bottomrule
		 	\end{tabular}\egroup
		 	\caption{\label{tab:funcvaluesn3} Different values for $x \in \mathbb{F}_3^3$, $f_3(x)$, $2 \cdot (x_0 \oplus x_1 \oplus x_2)$ and the expectation value of the respective measurement operators with respect to the four-qutrit GHZ state, which evaluates to	$\alpha^{\sum_{i=0}^{2} x_i + 2 \cdot  (x_0\oplus x_1 \oplus x_2)}$ for the computation of $f_3(x)$ as defined in Eqn.~\eqref{eq:func}. As $f_3(x)$ is invariant under coordinate permutation, we only need to check for different combinations of $x_0$, $x_1$ and $x_2$, where $x_0, x_1, x_2 \in \GFt$.}
 \end{table*}

 Next, we perform the induction step $n \mapsto n+1$:

\begin{enumerate}\setcounter{enumi}{1}
	\item $n \mapsto n+1$. It is not difficult to separate the expression on the l.h.s. of Eqn.~\eqref{eq:indhyp} into a part with all $x_i$ except $x_n$ and one with $x_n$. It is the r.h.s. that is slightly more challenging. To separate the exponent in Eqn.~\eqref{eq:indhyp} into a part containing $x_n$ and one without it, observe that:
 
	\begin{equation}\label{eq:fnfnplus1}
	f_{n+1}(x) = f_{n}(x)  \oplus x_n^2 \oplus 2 \cdot \bigoplus_{i=0}^{n-1} x_n^2 x_i \oplus x_n \left[ \bigoplus_{i=0}^{n-1}x_i \oplus 2 \cdot \bigoplus_{i=0}^{n-1}x_i^2 \oplus \bigoplus_{i=0}^{n-2} \bigoplus_{j=i+1}^{n-1}x_ix_j\right].
	\end{equation}

One arrives at that conclusion by splitting up the sum over all $x_i^2$ at $i=n$ and by factoring out the $x_n^2$ and $x_n$ in all terms that contain it. The terms that do not contain any non-zero power of $x_n$ are part of $f_n(x)$. One can verify that the expression in square brackets in Eqn.~\eqref{eq:fnfnplus1} is equal to:

\begin{equation} \label{eq:squarebrackfactor}
\left(2 \bigoplus_{i=0}^{n-1} x_i \right) \cdot \left(2 \oplus \bigoplus_{i=0}^{n-1} x_i \right).
\end{equation}

That is because multiplying these two expression leads to one term containing the sum over all $x_i$ (whereby the $2 \cdot 2$ becomes 1 due to mod 3 arithmetic), to one term containing the sum over all $x_i^2$ (so the 2 stays there) and to two terms containing the sum over all combinations of $x_i$ and $x_j,\, i \neq j$. (Both of these terms are preceded by a 2, resulting in a $2 \oplus 2 =1$). \newline
The two expressions in Eqn.~\eqref{eq:squarebrackfactor} only contain the sums (mod 3) over all $x_i$. We abbreviate $\bigoplus_{i=0}^{n-1} x_i$ by $S$ (for sum) and arrive at:
 
\begin{equation} \label{eq:fnplus1xP}
f_{n+1}(x) = f_{n}(x)  \oplus x_n^2 \oplus 2 \cdot x_n^2  \cdot S \oplus x_n \cdot \left(2 \cdot S\right) \cdot \left(2 \oplus S \right).
\end{equation}

We have obtained an expression for $f_{n+1}(x)$ that contains only $f_n(x)$, $x_n$ and $S$, the sum over all $x_i$ for $i<n$. Therefore, in order to check if the induction hypothesis Eqn.~\eqref{eq:indhyp} is fulfilled for $n+1$, it remains to differentiate between a handful of cases. We insert Eqn.~\eqref{eq:fnplus1xP} and $S = \bigoplus_{i=0}^{n-1} x_i$ into Eqn.~\eqref{eq:indhyp} and set $n \mapsto n+1$:

\begin{equation}
\alpha^{\sum_{i=0}^{n-1} (x_i)+x_n +2\cdot (S \oplus x_n)}  
= \omega^{f_n(x) \oplus x_n^2 \oplus 2 \cdot x_n^2 \cdot S \oplus x_n \cdot \left(2 \cdot S\right) \cdot \left(2 \oplus S \right)}  \, \quad \forall x.
\end{equation}

Now, we insert the induction hypothesis for $n$:
 
	\begin{align}
	 \alpha^{\sum_{i=0}^{n-1} (x_i)+x_n +2\cdot (S \oplus x_n)} &= \alpha^{\sum_{i=0}^{n-1}(x_i) + 2 \cdot S} \cdot \omega^{x_n^2 \oplus 2 \cdot x_n^2 \cdot S \oplus x_n \cdot  (2 \cdot S) \cdot (2 \oplus S)}  \, \quad \forall x  \\
	 	\implies  \alpha^{x_n +2\cdot (S \oplus x_n)} &=  \alpha^{2\cdot S} \cdot \omega^{x_n^2 \oplus 2 \cdot x_n^2 \cdot S \oplus x_n \cdot \left(2 \cdot S\right) \cdot \left(2 \oplus S \right)}  \, \quad \forall x. \label{eq:indhypPxn} 
	\end{align}

Equation Eqn.~\eqref{eq:indhypPxn} is a condition that must hold for Proposition \ref{prop:exfunc} to be true and only contains $x_n$ and $S$.
With this, we can check for every combination of $x_n$ and $S$.
\begin{enumerate}
	
\item	$x_n = 0$, $S \in \{0,1,2\}$. In this case, Eqn.~\eqref{eq:indhypPxn} is trivially fulfilled.
\item $x_n =1$, $S=0$. Here, the l.h.s. of Eqn.~\eqref{eq:indhypPxn} yields $\alpha^{1 +2} = \omega^1$ and the r.h.s. results in: 

\begin{equation}
 \omega^{1^2 \oplus 2\cdot 1 \cdot 0  \oplus 1 \cdot \left(2 \cdot 0\right) \cdot \left(2 \oplus 0\right)} = \omega^1.
\end{equation} 
\item $x_n =2$, $S=0$. The l.h.s. of Eqn.~\eqref{eq:indhypPxn} is $\alpha^{2 +1} = \omega^1$ and the r.h.s. results in: 

\begin{equation}
\omega^{2^2 \oplus 2\cdot 1 \cdot 0  \oplus 2 \cdot \left(2 \cdot 0\right) \cdot \left(2 \oplus 0\right)} = \omega^{2^2}=\omega^1.
\end{equation} 

\item $x_n =1$, $S=1$. The l.h.s. of Eqn.~\eqref{eq:indhypPxn} is $\alpha^{1 +2 \cdot 2} = \alpha^{2}$ and the r.h.s. results in: 

\begin{equation}
\alpha^2 \cdot \omega^{1 \oplus 2\cdot 1 \cdot 1  \oplus 1 \cdot \left(2 \cdot 1\right) \cdot \left(2 \oplus 1\right)} = \alpha^2  \omega^{0}.
\end{equation} 

\end{enumerate}

\end{enumerate}

Analogously, one can show that Eqn.~\eqref{eq:indhypPxn} holds for all other combinations of $x_n$ and $S$. This completes the proof of Proposition \ref{prop:exfunc}. \qed

%
\section{Proof of Proposition \ref{prop:ScalingConstant}}\label{sec:AppPropExp}
%

To prove Proposition \ref{prop:ScalingConstant}, we first show that one can compute a function with a distance of 1 to a constant one with 3-NMQC and a pre-processing matrix $P$ if and only if one can compute it with a pre-processing matrix $P'=PM$, where $M$ is an $n$ by $n$ ternary invertible matrix. Then we show that if every row in $P$ does not contain all 1s or 2s, it is impossible to compute the function. In the end we combine these two facts and show that, given $l < (3^n-1)/2$, one can always find an invertible matrix $M$ that turns $P$ into a $P'$ in which every row is unequal to $(1,1,1,..,1)$ and $(2,2,2,...,2)$, which completes the proof.

We begin by, without loss of generality, fixing the function $f(x)$ to be computed as:

\begin{equation} \label{eq:FuncHamDist1toConst}
	f(x) = \begin{cases}
		0 & x= \vec{0} \\
		1 & \text{else,}
	\end{cases}
\end{equation}
so $f(x)$ differs from the constant function $g(x)=1$ at $\tilde{x}=\vec{0}$. Other cases can be retrieved from this one by relabelling in- and output bits. Now, if we apply condition Eqn.~\eqref{eq:ConditionSuccess1}, we arrive at:
\begin{equation}  \label{eq:ConditionSuccessProof}
	\alpha^{\sum_i \left(\sum_j (P_{ij} x_j)_{\oplus} \right)  \phi_i} = \omega^{f(x)+c}.
\end{equation}

The first step is to observe that Eqn.~\eqref{eq:ConditionSuccessProof}, for this particular $f(x)$, is equivalent to:
\begin{equation}  \label{eq:ConditionSuccessProofM}
	\alpha^{\sum_i \left(\sum_j (P'_{ij} x_j)_{\oplus} \right)  \phi_i} = \omega^{f(x)+c},
\end{equation}
where $P'=PM$. (As a reminder, $\alpha = e^{\frac{2\pi i}{9}}$.) That is because $P'=PM$ leaves the $x=\vec{0}$ case unchanged (as in this case one multiplies the row $P_i$ by a zero-vector and the only non-zero term left to be chosen is the post-processing trit $c$.) All other cases lead to the same term on the r.h.s, namely $\omega^{1+c}$. So if Eqn.~\eqref{eq:ConditionSuccessProof} is fulfilled for all $x$, then Eqn.~\eqref{eq:ConditionSuccessProofM} is too. This, in turn means that, if we find a $P'=PM$ for which Eqn.~\eqref{eq:ConditionSuccessProofM} is not fulfilled, Eqn.~\eqref{eq:ConditionSuccessProof} does not hold. This was the first step of the proof.

We continue with the second step, showing that Eqn.~\eqref{eq:ConditionSuccessProof} leads to a contradiction if every row of $P$ is unequal to a row of all 1s or a row of all 2s. If we apply condition Eqn.~\eqref{eq:ConditionSuccessProof} for $x=\vec{0}$, we get:
\begin{equation} 
	\alpha^{0} = \omega^{f(\vec{0})+c}  =  \omega^c.
\end{equation}

We conclude that $c$ must be zero. For $x \neq \vec{0}$, one arrives at:
\begin{equation}  \label{eq:ProofConstantAlphaCond}
	\alpha^{\sum_i \left(\sum_j (P_{ij} x_j)_{\oplus} \right)  \phi_i} = \omega^{1} \implies 	\sum_i \left(\sum_j (P_{ij} x_j)_{\oplus} \right)  \phi_i  = 3+9k_x,
\end{equation}
where $k_x$ is an integer dependent on $x$.  

Now we introduce the sum:
	\begin{equation}\label{eq:ProofConstantSum}
		\sum_{x \neq 0} \omega^{\sum_k x_k}  \sum_i \left(\sum_j (P_{ij} x_j)_{\oplus} \right)  \phi_i =  \sum_i  \sum_{x \neq 0} \left( \omega^{\sum_k x_k}  \sum_j (P_{ij} x_j)_{\oplus} \right)  \phi_i. 
	\end{equation}

The r.h.s. of Eqn.~\eqref{eq:ProofConstantSum} is zero if every row $P_i$ in $P$ contains at least one $P_{ij} = 0$. One realises that by looking at the sum on the r.h.s. over $x$:
\begin{equation} \label{eq:ProofConstantSumX} 
	\sum_{x \neq 0} \left( \omega^{\sum_k x_k}  \sum_j (P_{ij} x_j)_{\oplus} \right).
\end{equation}

Without loss of generality, we fix $P_{i,n-1} =0$. As the last element of the row $P_i$ is zero, the last element of $x$ does not matter and if all elements before that one are zero, the sum over the elements of $P_i$ is zero too:
\begin{equation}  
	\sum_j (P_{ij} x_j)_{\oplus} = 0 \qquad \text{for } x = (0,...,0,d)^T,d\in\{1,2\}.
\end{equation}

We are left with $3^n-3$ presumably non-zero terms in the sum Eqn.~\eqref{eq:ProofConstantSumX}. We can write $x$ as $x = (x',x_{n-1})^T$, where $x'$ is an $n-1$ dimensional vector with entries in $\GFt$. As the inner sum in Eqn.~\eqref{eq:ProofConstantSumX} is the same for $x = (x',0)^T$, $x = (x',1)^T$ and $x = (x',2)^T$, the sum of the terms in Eqn.~\eqref{eq:ProofConstantSumX} over these three different values of $x$ is:
	\begin{align}  
		\begin{split}
			&\omega^{\sum_k x'_k}  \sum_j (P_{ij} x'_j)_{\oplus}  +   \omega^{\sum_k (x'_k)+1}  \sum_j (P_{ij} x'_j)_{\oplus}  +\omega^{\sum_k (x'_k)+2}  \sum_j (P_{ij} x'_j)_{\oplus}\\
			&= 
			\left(\omega^{\sum_k (x'_k)}  \sum_j (P_{ij} x'_j)_{\oplus}\right) \left(\omega^0 + \omega^1 + \omega^2\right)=0.
			\end{split}
	\end{align}
(The sum over the roots of one is zero.) As we can divide all $3^n-3$ left over $x$ into $3^{n-1}-1$ sets $\{(x',0)^T,(x',1)^T,(x',2)^T\}$ for $x' \in F_q^{n-1}$, one sees that Eqn.~\eqref{eq:ProofConstantSumX} is zero, if there is a zero in every row $P_i$ of $P$:
	\begin{equation} \label{eq:ProofConstantSum0forP0}
		\sum_{x \neq 0} \left( \omega^{\sum_k x_k}  \sum_j (P_{ij} x_j)_{\oplus} \right) = 0 \qquad \text{if } \forall i \in \{0,...,l-1\} \exists j \in \{0,...,n-1\}: P_{ij} =0
	\end{equation}
 
This is not the only case in which the sum in Eqn.~\eqref{eq:ProofConstantSum} is zero. In fact:
	\begin{equation} \label{eq:ProofConstantSum0forPnot111}
		\sum_{x \neq 0} \left( \omega^{\sum_k x_k}  \sum_j (P_{ij} x_j)_{\oplus} \right) = 0
		 \qquad \forall P_i \neq(c,c,c,...,c), \, c\in \{1,2\}.
	\end{equation}

So the sum Eqn.~\eqref{eq:ProofConstantSumX} is zero if $P_i$ is not a row containing only 1s or a row containing only 2s.  
This is due to the fact that we can split the sum into different terms (for $P_{ij} \neq 0 \forall j$):
	\begin{align}
		\begin{split}
		&\sum_{x \neq 0} \left( \omega^{\sum_k x_k}  \sum_j (P_{ij} x_j)_{\oplus} \right) \\
		&= 	\sum_{\substack{x \neq 0 \\ P_{ij} x_j   = 0}} \left( \omega^{\sum_k x_k}  \sum_j (P_{ij} x_j)_{\oplus} \right) 
		+ \sum_{P_{ij} x_j =1} \left( \omega^{\sum_k x_k}  \sum_j (P_{ij} x_j)_{\oplus} \right)  +\sum_{P_{ij} x_j = 2} \left( \omega^{\sum_k x_k}  \sum_j (P_{ij} x_j)_{\oplus} \right).
		\end{split}
	\end{align}

As the order of the index $j$ does not matter, we reorder $P_{ij}$ and $x_j$ such that $P_{ij} = 1$ for $j = 0,..,m-1$ and $P_{ij} = 2$ for $j\geq m$. ($P_{ij} \neq 0 \quad \forall i, j$ as we have already covered this case above.) This results in:
	\begin{align} 
		\begin{split}
			&\sum_{x \neq 0} \left( \omega^{\sum_k x_k}  \sum_j (P_{ij} x_j)_{\oplus} \right) = 	\sum_{\substack{x \neq 0 \\ P_{ij} x_j   = 0}} \left( \omega^{\sum_k x_k}  \sum_j (P_{ij} x_j)_{\oplus} \right) \\
			&\phantom{=}+ \sum_{P_{ij} x_j =1} \left( \omega^{\sum_k x_k}  \sum_j (P_{ij} x_j)_{\oplus} \right) 
			+\sum_{P_{ij} x_j = 2} \left( \omega^{\sum_k x_k}  \sum_j (P_{ij} x_j)_{\oplus} \right) \\
			& =\sum_{\substack{x \neq 0 \\ \sum_{j=0}^{m-1} x_j   + 2\sum_{j\geq m}  x_j}=0} \left( \omega^{\sum_k x_k}  \cdot 0 \right) 
			+\sum_{\substack{x \neq 0 \\ \sum_{j=0}^{m-1} x_j   + 2\sum_{j\geq m}  x_j }=1} \left( \omega^{\sum_k x_k}  \cdot 1 \right) \\
			&\phantom{=}+\sum_{\substack{x \neq 0 \\ \sum_{j=0}^{m-1} x_j   + 2\sum_{j\geq m} x_j }=2} \left( \omega^{\sum_k x_k}   \cdot 2\right) \\
			& =\sum_{\substack{x \neq 0 \\ \sum_{j=0}^{m-1} x_j = 1   \\ \sum_{j\geq m} x_j=0}}   \omega^{\sum_k x_k} + \sum_{\substack{x \neq 0 \\ \sum_{j=0}^{m-1} x_j = 2   \\ \sum_{j\geq m}  x_j=1}}   \omega^{\sum_k x_k} + \sum_{\substack{x \neq 0 \\ \sum_{j=0}^{m-1} x_j = 0   \\ \sum_{j\geq m}  x_j=2}}   \omega^{\sum_k x_k}  \\
			&\phantom{=}+ \sum_{\substack{x \neq 0 \\ \sum_{j=0}^{m-1} x_j = 2   \\ \sum_{j\geq m} x_j=0}}   2\omega^{\sum_k x_k} + \sum_{\substack{x \neq 0 \\ \sum_{j=0}^{m-1} x_j = 1   \\ \sum_{j\geq m}  x_j=2}} 2  \omega^{\sum_k x_k} + \sum_{\substack{x \neq 0 \\ \sum_{j=0}^{m-1} x_j = 0   \\ \sum_{j\geq m+1}  x_j=1}} 2  \omega^{\sum_k x_k}.
		\end{split}
	\end{align}
 
There are $3^{m-1}$  $x = (x_0,...,x_{m-1})$ such that $\sum_{j=0}^{m-1} x_j = l$ and $3^{n-m-1}$ $x = (x_m,...,x_{n-1})$ such that $\sum_{j=m}^{n-1} x_j = l$ for any $l \in \GFt$. We continue:
	\begin{align}
		\begin{split}
			&\sum_{x \neq 0} \left( \omega^{\sum_k x_k}  \sum_j (P_{ij} x_j)_{\oplus} \right) =   3^{m-1} \cdot 3^{n-m-1}\omega^{1} + 3^{m-1} \cdot 3^{n-m-1}  \omega^{0} + 3^{m-1} \cdot 3^{n-m-1}   \omega^{2}   \\ 
			&\phantom{=} + 3^{m-1} \cdot 3^{n-m-1}\cdot  2\omega^{2}  + 3^{m-1} \cdot 3^{n-m-1} \cdot 2  \omega^{0} +3^{m-1} \cdot 3^{n-m-1} \cdot 2  \omega^{1} \\
			&= 3^{n-2} (\omega+\omega^1+\omega^2) +2 \cdot 3^{n-2} (\omega+\omega^1+\omega^2) \\
			& =0.
		\end{split}
	\end{align}

Plugging Eqn.~\eqref{eq:ProofConstantSum0forPnot111} and Eqn.~\eqref{eq:ProofConstantSum0forP0} into Eqn.~\eqref{eq:ProofConstantAlphaCond}, one arrives at:
\begin{equation} \label{eq:ProofConstantSumXRHS}
	\sum_{x \neq 0} \omega^{\sum_k x_k} (3+9k_x) = 0.
\end{equation}
In the first sum $\sum_{x\neq 0} 3 \omega^{\sum_k x_k}$ all terms except the ones for $x = (0,..,1)^T$ and $x = (0,...,2)^T$ cancel out due to the same reasons as above and we are left with:
\begin{equation}
	\sum_{x \neq 0} \omega^{\sum_k x_k} 3= 3 (\omega^1 + \omega^2)=-3 .
\end{equation}
The second sum $\sum_{x\neq 0} 9 \omega^{\sum_k x_k} k_x$ can be split up into three terms:
\begin{equation} \label{eq:ProofConstantSum9om}
	\sum_{x \neq 0} 9 k_x  \omega^{\sum_k x_k} = 9 (K_0 +K_1+K_2),
\end{equation}
where $K_m = \sum_{\substack{x \neq 0, \\ \sum_k x_k =m}} \omega^m  k_x$. 
Clearly, $K_0$ is an integer (as $\omega^0 =1$). However, $K_1$ and $K_2$ are multiples of $\omega^1$ and $\omega^2$ respectively and therefore complex numbers. Inserting Eqn.~\eqref{eq:ProofConstantSum9om} into equation Eqn.~\eqref{eq:ProofConstantSumXRHS}, one arrives at:
\begin{equation}\label{eq:ProofConstantSumXRHS2}
		-3 + 9(K_0+K_1+K_2) 
		= -1 + 9 (\Im{K_1} + \Im{K_2})  
		+9 (\Re{K_1}+\Re{K_2} +K_0) =0.
\end{equation}
The two imaginary parts $\Im{K_1}$ and $\Im{K_2}$ in Eqn.~\eqref{eq:ProofConstantSumXRHS2} are multiples of $\Im{\omega^1}$ and $\Im{\omega^2}$:
\begin{equation}
	\Im{K_1}  = \tilde{K_1} \Im{\omega}, \quad \Im{K_2}  = \tilde{K_2} \Im{\omega^2},
\end{equation}
where $\tilde{K_1}$ and $\tilde{K_2}$ are integers. The two real parts $\Re{K_1}$ and $\Re{K_2}$ are multiples of $\Re{\omega^1}$ and $\Re{\omega^2}$ with the same factors $\tilde{K_1}$ and $\tilde{K_2} $.

As $\Im{\omega} = -\Im{\omega^2}$ and the total imaginary part on the left hand side in Eqn.~\eqref{eq:ProofConstantSumXRHS2} must be zero, $\tilde{K_1}$ must be equal to $\tilde{K_2} \equiv \tilde{K}$, hence:
\begin{align} \label{eq:Proofmin3pl9zero}
	\begin{split}
		-3 + 9(K_0+K_1+K_2) 
		&= -3 + 9 (\Im{K_1} + \Im{K_2}) +9 (\Re{K_1}+\Re{K_2} +K_0)  \\
		&=	-3 + 9 \cdot 0 +9 (\tilde{K_1} \Re{\omega} +\tilde{K_2} \Re{\omega}+K_0)  \\
		&= -3  +9 (-\frac{1}{2}\tilde{K}   -\frac{1}{2}\tilde{K}  +K_0) \\
		&= -3  +9 (- \tilde{K}  +K_0)  \\
		& = 0.
	\end{split}
\end{align}
In the third line of Eqn.~\eqref{eq:Proofmin3pl9zero} we inserted $\Re{\omega}= \Re{\omega^2}=\Re{e^{\frac{2\pi i}{3}}}= \Re{e^{\frac{4\pi i}{3}}}=-\frac{1}{2}$.

Evidently, this cannot be fulfilled if $\tilde{K}$ and $K_0$ are integers. Therefore, condition Eqn.~\eqref{eq:ProofConstantAlphaCond} leads to a contradiction and cannot be fulfilled if $P_{ij} \neq (c,c,...,c), \, c\in \{1,2\} $.  The second step is complete.\\

It remains to show that, given $l< (3^n-1)/2$, one can always transform $P$ into a $P'$ that leads to such a contradiction. If we have less than $3^n-1$ qutrits, the rows of $P$ will not contain every possible combination for a string of $n$ trits. We denote one missing row by $y$. We can always find an invertible (hence injective) map $M: \GFtn \mapsto \GFtn$ that maps  $y' \equiv (1,1,...,1)$ to the missing row $y$

\begin{equation} \label{eq:ProofMcond}
	y' M = y.
\end{equation}

Conversely, the inverse $M^{-1}$ maps all strings $\tilde{y} \neq y$ to strings $\tilde{y'} \neq y'$. This, alongside the fact that condition Eqn.~\eqref{eq:ProofConstantAlphaCond} is equivalent to: 

\begin{equation}  \label{eq:ProofConstantAlphaCond2}
	\sum_i \left(\sum_j (P'_{ij} x_j)_{\oplus} \right)  \phi_i  = 3+9k_x,
\end{equation}
where $P' = PM$ for some invertible map $M: \GFtn \mapsto \GFtn$, would finalise the proof, as we could always find an $M$ such that Eqn.~\eqref{eq:ProofMcond} holds, leading to the contradiction derived above.  However, the sum in Eqn.~\eqref{eq:ProofConstantSum0forPnot111} is also zero for $P_i = (2,...,2)$. This is why we need no fewer than $(3^n-1)/2$ qutrits:  
If $l < (3^n-1)/2$, this means that there are at least $(3^n-1)/2+1$ non-trivial missing strings. This means that we can find a missing string $y$ whose additive inverse $2y$ is also missing. We can then find an $M$ that fulfils Eqn.~\eqref{eq:ProofMcond}, and it will automatically fulfil:
\begin{equation}
	(2y') M = 2y,
\end{equation}
where $y'$ is a string containing all ones. Therefore, the inverse $M^{-1}$ maps all strings $ \tilde{y} \neq y, 2y$ to anything but a string containing all ones or all twos. The third step is complete, and hence the proof. \qed
\newpage
 \end{widetext}

\end{document}